\documentclass[aip,amsmath,amssymb,nofootinbib,preprint]{revtex4-2}

\usepackage{dcolumn}
\usepackage[latin1,utf8]{inputenc}
\usepackage[T1]{fontenc}
\usepackage{amsmath}
\usepackage{mathptmx}
\usepackage{amsmath,amsfonts,mathtools,amssymb}
\usepackage{float}
\usepackage[dvipsnames]{xcolor}
\usepackage{bm,nicefrac}
\usepackage{graphicx}
\usepackage{epsfig}
\usepackage{epstopdf}

\usepackage{tabularx}
\usepackage{subcaption}
\usepackage[section]{placeins}
\usepackage{textgreek}

\usepackage[toc,page]{appendix}

\usepackage{cancel}

\definecolor{my-gray}{gray}{0.95}
\definecolor{my-green}{HTML}{15B01A}
\definecolor{my-purple}{HTML}{9A0EEA}
\definecolor{my-yellow}{HTML}{FAC205}
\usepackage{listings}

\usepackage[breaklinks=False,colorlinks=true,linkcolor=blue,urlcolor=blue,citecolor=blue]{hyperref} 

\begin{document}
\newcommand{\covBpartial}[2]{\ensuremath{\partial_{#1}B_{#2} &=
\left(\partial_{#1}B^u\eu + B^u\mathbf{e}_{u#1} + \partial_{#1}B^v\ev + B^v\mathbf{e}_{v#1}\right)\cdot \mathbf{e}_{#2} \\&+ \left(B^u\eu + B^v\ev \right)\cdot \mathbf{e}_{#2 #1} }}

\newcommand{\iotabar}{\mbox{$\,\iota\!\!$-}}
\newcommand{\bB}{\bm{B}}
\newcommand{\Rs}{\partial_{s}R}
\newcommand{\Zs}{\partial_{s}Z}
\newcommand{\Rv}{\partial_{v}R}
\newcommand{\Zv}{\partial_{v}Z}
\newcommand{\Ru}{\partial_{u}R}
\newcommand{\Zu}{\partial_{u}Z}
\newcommand{\Rss}{\partial_{ss}R}
\newcommand{\Zss}{\partial_{ss}Z}

\newcommand{\sg}{\sqrt{g}}
\newcommand{\gs}{\partial_s\sqrt{g}}
\newcommand{\gu}{\partial_u\sqrt{g}}
\newcommand{\gv}{\partial_v\sqrt{g}}
\newcommand{\gss}{\partial_{ss}\sqrt{g}}
\newcommand{\gsu}{\partial_{su}\sqrt{g}}
\newcommand{\gsv}{\partial_{sv}\sqrt{g}}
\newcommand{\gsss}{\partial_{sss}\sqrt{g}}
\newcommand{\gssu}{\partial_{ssu}\sqrt{g}}
\newcommand{\gssv}{\partial_{ssv}\sqrt{g}}

\newcommand{\lv}{\frac{\partial\lambda}{\partial v}}
\newcommand{\lu}{\frac{\partial\lambda}{\partial u}}

\newcommand{\lsv}{\frac{\partial^2\lambda}{\partial s \partial v }}
\newcommand{\luv}{\frac{\partial^2\lambda}{\partial u \partial v }}
\newcommand{\lsu}{\frac{\partial^2\lambda}{\partial s \partial u }}

\newcommand{\luu}{\frac{\partial^2\lambda}{\partial u^2 }}
\newcommand{\lss}{\frac{\partial^2\lambda}{\partial s^2 }}
\newcommand{\lvv}{\frac{\partial^2\lambda}{\partial v^2 }}

\newcommand{\lssu}{\frac{\partial^3\lambda}{\partial s^2 \partial u}}
\newcommand{\lssv}{\frac{\partial^3\lambda}{\partial s^2 \partial v}}
\newcommand{\lsuv}{\frac{\partial^3\lambda}{\partial s \partial u \partial v}}
\newcommand{\lsvv}{\frac{\partial^3\lambda}{\partial s \partial v^2}}
\newcommand{\lsuu}{\frac{\partial^3\lambda}{\partial s \partial u^2}}

\newcommand{\Rrr}{\partial_{\rho\rho}R}
\newcommand{\Zrr}{\partial_{\rho\rho}Z}
\newcommand{\Rvv}{\partial_{\vartheta\vartheta}R}
\newcommand{\Zvv}{\partial_{\vartheta\vartheta}Z}
\newcommand{\Rrv}{\partial_{\rho\vartheta}R}
\newcommand{\Zrv}{\partial_{\rho\vartheta}Z}
\newcommand{\Rzr}{\partial_{\zeta\rho}R}
\newcommand{\Zzr}{\partial_{\zeta\rho}Z}
\newcommand{\Rzv}{\partial_{\zeta\vartheta}R}
\newcommand{\Zzv}{\partial_{\zeta\vartheta}Z}
\newcommand{\Rzz}{\partial_{\zeta\zeta}R}
\newcommand{\Zzz}{\partial_{\zeta\zeta}Z}
\newcommand{\Rrrv}{\partial_{\rho\rho\vartheta}R}
\newcommand{\Zrrv}{\partial_{\rho\rho\vartheta}Z}
\newcommand{\Rrvv}{\partial_{\rho\vartheta\vartheta}R}
\newcommand{\Zrvv}{\partial_{\rho\vartheta\vartheta}Z}
\newcommand{\Rzrv}{\partial_{\zeta\rho\vartheta}R}
\newcommand{\Zzrv}{\partial_{\zeta\rho\vartheta}Z}
\newcommand{\Rrrvv}{\partial_{\rho\rho\vartheta\vartheta}R}
\newcommand{\Zrrvv}{\partial_{\rho\rho\vartheta\vartheta}Z}

\newcommand{\lrh}{\frac{\partial\lambda}{\partial \rho}}
\newcommand{\lze}{\frac{\partial\lambda}{\partial \zeta}}
\newcommand{\lth}{\frac{\partial\lambda}{\partial \theta}}

\newcommand{\eS}{{\mathbf e}^{s}}
\newcommand{\eV}{{\mathbf e}^{v}}
\newcommand{\eU}{{\mathbf e}^{u}}

\newcommand{\eR}{{\mathbf e}^{\rho}}
\newcommand{\eT}{{\mathbf e}^{\theta}}
\newcommand{\eZ}{{\mathbf e}^{\zeta}}

\newcommand{\es}{{\mathbf e}_{s}}
\newcommand{\eu}{{\mathbf e}_{u}}
\newcommand{\ev}{{\mathbf e}_{v}}
\newcommand{\ess}{{\mathbf e}_{ss}}
\newcommand{\esv}{{\mathbf e}_{sv}}
\newcommand{\esu}{{\mathbf e}_{su}}
\newcommand{\evs}{{\mathbf e}_{vs}}
\newcommand{\evv}{{\mathbf e}_{vv}}
\newcommand{\evu}{{\mathbf e}_{vu}}
\newcommand{\eus}{{\mathbf e}_{us}}
\newcommand{\euv}{{\mathbf e}_{uv}}
\newcommand{\euu}{{\mathbf e}_{uu}}
\newcommand{\evus}{{\mathbf e}_{vus}}
\newcommand{\evvs}{{\mathbf e}_{vvs}}
\newcommand{\euvs}{{\mathbf e}_{uvs}}
\newcommand{\esss}{{\mathbf e}_{sss}}
\newcommand{\euss}{{\mathbf e}_{uss}}
\newcommand{\eusss}{{\mathbf e}_{usss}}
\newcommand{\essu}{{\mathbf e}_{ssu}}
\newcommand{\eusu}{{\mathbf e}_{usu}}
\newcommand{\eussu}{{\mathbf e}_{ussu}}
\newcommand{\evsu}{{\mathbf e}_{vsu}}

\newcommand{\evss}{{\mathbf e}_{vss}}
\newcommand{\essv}{{\mathbf e}_{ssv}}
\newcommand{\evsv}{{\mathbf e}_{vsv}}
\newcommand{\eusv}{{\mathbf e}_{usv}}
\newcommand{\eussv}{{\mathbf e}_{ussv}}

\newcommand{\erh}{{\mathbf e}_{\rho}}
\renewcommand{\eth}{{\mathbf e}_{\theta}}
\newcommand{\eze}{{\mathbf e}_{\zeta}}

\newcommand{\erer}{\vert{\mathbf e}_{\rho}\vert^2}
\newcommand{\evev}{\vert{\mathbf e}_{\vartheta}\vert^2}
\newcommand{\ezez}{\vert{\mathbf e}_{\zeta}\vert^2}
\newcommand{\erev}{(\er\cdot\ev)}
\newcommand{\erez}{(\er\cdot\ez)}
\newcommand{\evez}{(\ev\cdot\ez)}
\newcommand{\erevz}{(\er\cdot\evz)}
\newcommand{\etez}{(\et\cdot\ez)}
\newcommand{\etev}{(\et\cdot\ev)}
\newcommand{\etet}{\vert{\mathbf e}_{\theta}\vert^2}
\newcommand{\ervez}{(\erv\cdot\ez)}
\newcommand{\evezr}{(\ev\cdot\ezr)}

\title{The DESC Stellarator Code Suite Part I: Quick and accurate equilibria computations}
\author{D. Panici}
\email[]{dpanici@princeton.edu}
\affiliation{Princeton University, Princeton, New Jersey 08544}

\author{R. Conlin}
\email[]{wconlin@princeton.edu}
\affiliation{Princeton University, Princeton, New Jersey 08544}

\author{D. W. Dudt}
\email[]{ddudt@princeton.edu}
\affiliation{Princeton University, Princeton, New Jersey 08544}

\author{K. Unalmis}
\email[]{kunalmis@princeton.edu}
\affiliation{Princeton University, Princeton, New Jersey 08544}

\author{E. Kolemen}
\email[]{ekolemen@princeton.edu}
\affiliation{Princeton University, Princeton, New Jersey 08544}

\begin{abstract}
3D equilibrium codes are vital for stellarator design and operation, and high-accuracy equilibria are also necessary for stability studies. This paper details comparisons of two three-dimensional equilibrium codes, VMEC, which uses a steepest-descent algorithm to reach a minimum-energy plasma state, and DESC, which minimizes the magnetohydrodynamic (MHD) force error in real space directly. Accuracy as measured by satisfaction of MHD force balance is presented for each code, along with the computation time. It is shown that DESC is able to achieve more accurate solutions, especially near-axis. The importance of higher accuracy equilibria is shown in DESC's better agreement of stability metrics with asymptotic formulae. DESC's global Fourier-Zernike basis also yields the solution with analytic derivatives explicitly everywhere in the plasma volume, provides improved accuracy in the radial direction versus conventional finite differences, and allows for exponential convergence. Further, DESC can compute the same accuracy solution as VMEC in an order of magnitude less time. 
\end{abstract}

\maketitle

\section{Introduction}
In the design of any fusion device, the preliminary step is the computation of a plasma equilibrium state with the desired geometry. A plasma in an equilibrium state can be described by the ideal magnetohydrodynamic (MHD) equilibrium model:

\begin{subequations}\label{eq:mhd_eq}
\begin{equation}\label{eq:JxB}
    \bm{J}\times \bm{B} = \nabla p
\end{equation}
\begin{equation}\label{eq:amperes}
    \nabla \times \bm{B} = \mu_0 \bm{J}
\end{equation}
\begin{equation}\label{eq:gauss}
    \nabla \cdot \bm{B} = 0
\end{equation}
\end{subequations}

where $\bm{B}$ is the magnetic field, $\bm{J}$ is the current density, $p$ is the scalar pressure, and $\mu_0$ is the permeability of free space. 
The satisfaction of these equations implies that the plasma is in perfect force balance, i.e.
\begin{equation}\label{eq:F}
    \bm{F} = \bm{J}\times \bm{B} - \nabla p = 0
\end{equation}

everywhere in the plasma, and the plasma state also coincides with a stationary state in the plasma potential energy, 

\begin{equation}\label{eq:W}
    W = \int_V \left(\frac{B^2}{2\mu_0} + \frac{p}{\gamma -1} dV \right)
\end{equation}
where $V$ is the plasma volume and $\gamma$ is the adiabatic index.

In tokamaks, the plasma is typically taken to be axisymmetric, allowing the MHD equilibrium to be described by the Grad-Shafranov equation, for which exist analytic solutions \citep{cerfon_one_2010,guazzotto_simple_2021}, and efficient codes to numerically solve for equilibria \citep{lao_reconstruction_1985}. However, the problem becomes much more difficult without the assumption of axisymmetry, making stellarator equilibria more challenging to compute. Adding to the challenge are the singular currents predicted by ideal MHD to form at rational surfaces in 3D geometries \cite{helander_theory_2014}. While this is an important topic to note, it will not be elaborated upon in this work and is left to future endeavors.

Very few analytical solutions to the general 3D equilibrium problem are known \citep{rosenbluth_nonlinear_1973}, and so typically three-dimensional (3D) equilibria must be found numerically. Thus, a fast, robust, and accurate 3D equilibrium solver is necessary for stellarator optimization studies.  The current workhorse code for 3D equilibria is VMEC\citep{hirshman_steepestdescent_1983}, which is integrated into all current stellarator optimization workflows \citep{spong1998stellopt,stellopt,rose2019,simsopt2021}. While a relatively robust and widely-used code, VMEC still suffers from shortcomings stemming from its issues at the axis and its radial discretization, as well as its legacy design. A new 3D stellarator equilibrium code, DESC \citep{dudt_desc_2020,Dudt_DESC}, has been developed which can overcome these issues.\\

In this first part of a three-part series on DESC, a comparison of DESC and VMEC equilibria will be conducted to show the advantages of DESC's equilibrium solver. Section \ref{LitRev} will review the existing 3D equilibrium codes, while Section \ref{Codes} will detail the two codes compared in this paper, VMEC and DESC. Section \ref{methods} will define the method of comparison and accuracy metrics used, and Section \ref{results} presents the results of the comparison of accuracy in terms of equilibrium solution and stability calculations. \\

Part II \citep{conlin2022desc} of the three-part series presents a novel perturbation and continuation method used by the DESC code for solving and optimizing stellarator equilibria. The efficiency and utility of the method is shown in the computation of complicated equilibria, and highlights the benefits of automatic differentiation. Part III \citep{dudt2022desc} presents DESC's unique stellarator optimization capabilities made possible by the efficient equilibrium solver and the perturbation method described in the earlier parts, resulting in orders of magnitude speed-up in optimization. These advantages are shown in the context of quasi-symmetry optimization, where results are compared to conventional tools \citep{spong1998stellopt,stellopt}. Three different quasi-symmetry objective formulations are also shown, with the relative advantages of each compared, highlighting the flexibility of DESC as an optimization code.

\section{Literature Review}\label{LitRev}


\citet{kruskal_equilibrium_1958} first formulated solutions to the ideal MHD equilibrium problem as a variational principle, and showed that solutions to Eq. \eqref{eq:mhd_eq} are toroidal equilibria with nested flux surfaces and with pressure as a flux function.
The earliest 3D equilibrium codes utilized this principle, and discretized the spatial coordinates using finite difference schemes \citep{betancourt_equilibrium_1976}.  The BETA code used an inverse coordinate mapping and second-order finite differences motivated by the variational principle to minimize energy and calculate equilibria \citep{bauer_computational_1978}.
Later, \citet{chodura_3d_1981} found equilibria numerically by minimizing $W$ on an Eulerian cylindrical grid.\\

Eventually, spectral codes (using Fourier series representations in the poloidal and toroidal angles) were employed, which were shown to be substantially more efficient in calculating equilibria than pure difference methods. \citet{schwenn_fourier_1984} created FIT as a spectral upgrade of the TUBE equilibrium code. \citet{bhattacharjee_variational_1984} derived a variational method with a spectral Fourier series in angle and Hermite cubic B-splines in the radial direction, and used both a conventional inverse mapping and a mixed coordinate mapping. Hender's NEAR code \citep{hender_calculation_1985} used the same methodology as Chodura and Schluter, but replaced the cylindrical coordinate system with vacuum flux coordinates and Fourier-decomposed the problem in both angles. \citet{hirshman_steepestdescent_1983} detailed the VMEC code, which also solved the inverse equilibrium problem based on the variational principle and using poloidal and toroidal Fourier series. VMEC is widely used in the stellarator community for the improvement its formulation had over existing equilibrium codes, although the radial discretization can lead to inaccuracy near-axis, and will be discussed more in Section \ref{VMEC}. Additionally, an updated version of VMEC, GVEC is currently being developed \citep{banon_navarro_global_2020,hudson_free-boundary_2020}. DESC \citep{dudt_desc_2020,Dudt_DESC} is a recent pseudospectral code which employs a spectral Fourier-Zernike basis in all three coordinates, and finds equilibria by satisfying the MHD force balance Eq. \eqref{eq:JxB} directly at collocation nodes. This choice of spectral basis automatically satisfies the necessary constraints at the axis for analytic functions, and the code will be explained more in Section \ref{DESC}. Each of these codes assumes nested flux surfaces, so multiple magnetic axes (i.e. islands) cannot be represented in their equilibrium representation.\\

Other 3D equilibrium codes have been created which are able to handle islands and even chaotic regions. PIES \citep{reiman_calculation_1986} solves for the equilibrium magnetic field by iteratively evolving pressure-driven currents and re-solving for $\bm{B}$ with Ampere's Law, solving the differential equations by angular Fourier decomposition and finite differences for the radial discretization. The BETA code was rewritten as the spectral (in angles) BETAS code \citep{betancourt_betas_1988}, which used a coordinate system capable of representing non-nested flux surfaces, and later was the basis of the NSTAB \citep{taylor_high_1994} 3D equilibrium and stability code. NSTAB used a method of finding the magnetic axis location using a residue condition obtained from the variational principle, as opposed to constraining the axis location based on linear interpolation or Taylor expansion as done by previous codes. SIESTA \citep{hirshman_siesta_2011} is an iterative equilibrium solver, similar to PIES but based off of the energy principle, which can handle more complicated magnetic field topologies than BETAS, and relies on a VMEC solution for initialization of the solving procedure. The HINT2 code \citep{suzuki_development_2006-1}  solves the MHD equilibrium problem by introducing artificial viscosity and resistivity to the resistive MHD equations and relaxing to an equilibrium state on an Eulerian grid, without any assumption of nested flux surfaces. SPEC \citep{hudson_computation_2012} also uses a relaxation method, but in the MRxMHD framework, solving for equilibria using stepped, discontinuous pressure profiles. This method allows for very complicated magnetic field topology, but at an expense of requiring input profiles that may not be realistic. 
SPEC has recently implemented Zernike polynomials as their radial basis at the magnetic axis, which effectively handles the coordinate singularity present there, similar to DESC. \citep{hudson_free-boundary_2020,qu_coordinate_2020}

\section{Code Descriptions}\label{Codes}
\subsection{VMEC}\label{VMEC}

The most widely used 3D equilibrium code in the stellarator community at present is the Variational Moments Equilibrium Code (VMEC) \citep{hirshman_steepestdescent_1983}. VMEC constructs equilibria by minimizing the MHD energy \eqref{eq:W} through a variational principle. The base geometry is a cylindrical coordinate system $\mathbf{x} = (R,\phi,Z)$.
VMEC uses as its computational grid the coordinates $\bm{\alpha} = (s,u,v)$, with $s$ being a radial coordinate proportional to the normalized toroidal flux, $u$ a poloidal-like angle, and $v$ is the geometric toroidal angle (i.e. same as cylindrical $\phi$):

\begin{subequations}
\begin{equation}
    s = \frac{\psi}{\psi}_a \hspace{2mm}, \hspace{2mm} 0\leq s \leq 1
\end{equation}
\begin{equation}
    u = \theta^* - \lambda(s,u,v)\hspace{2mm}, \hspace{2mm} 0\leq u \leq 2\pi
\end{equation}
\begin{equation}
    v = \phi  \hspace{2mm}, \hspace{2mm} 0\leq v \leq 2\pi/N_{FP}
\end{equation}
\end{subequations}

where $\psi$ is the toroidal flux enclosed by a flux surface, normalized by $2\pi$, $\psi_a$ is the normalized toroidal flux enclosed by the plasma boundary (i.e. at s=1), $N_{FP}$ is the number of field periods in the configuration, and $\lambda(s,u,v)$ is a function periodic in $(u,v)$ that converts $u$ to a magnetic poloidal angle $\theta^*$ \citep{helander_theory_2014}.

VMEC solves the so-called inverse equilibrium problem, where the flux surface positions are taken to be functions of the computational coordinates, and the equilibrium is found by solving for the mappings $R = R(s,u,v), ~Z = Z(s,u,v)$ and the stream function $\lambda(s,u,v)$. These functions are expanded in a Fourier series in poloidal and toroidal angles as:

\begin{align}
\begin{split}
X(s,u,v) = \sum_{m=0}^M \sum_{n=-N}^{N} \big[X_{mn,c}(s)cos(mu - nvN_{FP}) + X_{mn,s}(s)sin(mu - nvN_{FP})\big]
\end{split}
\end{align}

where $X = \{R,Z,\lambda\}$. $R_{mn,c}(s)$ $,R_{mn,s}(s),Z_{mn,c}(s), Z_{mn,s}(s)$ are the Fourier coefficients of the flux surface at normalized toroidal flux $s$. The $c,s$ subscripts denote $cos$ and $sin$ coefficients, respectively. $m,n$ are the poloidal and toroidal mode numbers, $M,N$ are the poloidal and toroidal resolutions, with $0\leq m \leq M$ and $-N \leq n \leq N$. Many solutions of interest exhibit stellarator symmetry, that is, $R(s,-u,-v) = R(s,u,v),~Z(s,-u,-v) = -Z(s,u,v)$, and in these symmetric cases the $Z_{mn,c},~R_{mn,s},~\lambda_{mn,c}$ terms can be dropped from the representation, reducing the computational workload. With this Fourier decomposition, the spectral width is defined as \citep{hirshman_explicit_1998}:
\begin{equation}\label{SpecWidth}
    M(p,q) = \frac{\sum_m \sum_n m^{p+q} \left(R^2_{mn} + Z^2_{mn}\right)}{\sum_m \sum_n m^{p} \left(R^2_{mn} + Z^2_{mn}\right)}
\end{equation}
where $p \geq 0,~q>0$ and $R_{mn},~Z_{mn}$ are the Fourier coefficients for poloidal mode $m$ and toroidal mode $n$. $\lambda$ is chosen so as to create the most efficient Fourier representation of the surfaces, in the sense that it minimizes the spectral width \citep{hirshman_explicit_1998}. 
\\
Due to the spectral expansion being only in the angular coordinates, any radial derivatives necessary are calculated using first-order finite differences between neighboring flux surfaces.\\
Through Gauss' law and with the assumptions of nested flux surfaces ($\bB \cdot \nabla s = 0$) and pressure as a flux function($p = p(s)$), the magnetic field can be written in contravariant form as:
\begin{align}
    \bB &= \nabla s \times \nabla \theta^* + \nabla v \times \nabla \chi\\
    &= B^u \eu + B^v \ev
\end{align}

where $\chi(s)$ is the poloidal flux enclosed by the flux surface labelled $s$ normalized by $2\pi$. $\bm{e}_{\alpha_i} = \frac{\partial \bm{x}}{\partial \alpha_i}$ are the covariant basis vectors. The contravariant basis vectors are $\bm{e}^{\alpha_i} = \nabla \alpha_i$, and are related to the covariant basis by:
\begin{subequations}\label{eq:switch_and_duality}
\begin{align}
    \bm{e}_{\alpha_i} &= \frac{\bm{e}_{\alpha_j} \times \bm{e}_{\alpha_k}}{\sqrt{g}},~~i,j,k ~\text{cyc}~1,2,3\\
    \bm{e}_{\alpha_i} \cdot \bm{e}^{\alpha_j} &= \delta^{ij}
\end{align}
\end{subequations}

where the Jacobian $\sg$ is given by:
\begin{equation}\label{eq:jacobian}
    \sqrt{g} = \es\cdot\eu\times\ev = \left(\eS \cdot \eU \times \eV \right)^{-1}
\end{equation}
The contravariant components of the magnetic field are then:
\begin{subequations}
\begin{align}
    B^s &= 0, ~~\text{due to}~\bB\cdot \eS = 0\\
   B^u &= \frac{1}{\sqrt{g}} \Big(\chi' - \psi' \lv \Big)\\
B^v &= \frac{1}{\sqrt{g}} \psi'\Big(1 + \lu \Big)
\end{align}
\end{subequations}

where the prime denotes a radial derivative $\partial / \partial s$. Inserting this definition of $\bB$ into Eq. \eqref{eq:mhd_eq} yields:

\begin{equation}
    \bm{F} = F_{s} \nabla s + F_{\beta} \bm{\beta}
\end{equation}

with the two independent force components:

\begin{subequations}
\begin{align}
    F_{s} &= \sqrt{g} (J^{v} B^{u} - J^{u} B^{v}) + p'\\
    F_{\beta} &= J^{s}
\end{align}
\end{subequations}

and the vector $\bm{\beta}$ in the helical direction:

\begin{equation}
    \bm{\beta} = \sqrt{g} (B^v \nabla u - B^u \nabla v)
\end{equation}

The current density contravariant components are given as:

\begin{equation}\label{eq:J_VMEC}
    J^i =\bm{J} \cdot \nabla \alpha_i = \frac{\nabla \cdot (\bB \times \nabla \alpha_i)}{\mu_0}
\end{equation}

With these vector fields defined, VMEC then constructs a minimization scheme by taking the variation of the MHD energy in Eq. \eqref{eq:W}. This ultimately yields an equation for the variation of $W$ \citep{hirshman_steepestdescent_1983}:

\begin{equation}
    \frac{dW}{dt} = -\int F^{mn}_j \frac{\partial X_{mn,j}}{\partial t} dV
\end{equation}

Where $F^{mn}_j$ is the fourier components of the covariant force components $F_j = (F_R,F_{\lambda},F_Z)$ and $X_{mn,j}$ being the corresponding Fourier coefficients of $(R,\lambda,Z)$ . The direction of steepest descent is given by: 
\begin{equation}
    \frac{\partial X_{mn,j}}{\partial t} =  F^{mn}_j
\end{equation}

yielding the partial differential equations to be solved, as making $\frac{dW}{dt}=0$ means a minimum in energy, and an equilibrium configuration, has been found. In the VMEC code, the above time operator is replaced by a second-order Richardson scheme \citep{hirshman_steepestdescent_1983}:

\begin{equation}\label{eq:richardson}
    \frac{\partial^2 X_{mn,j}}{\partial t^2}+\frac{1}{\tau}\frac{\partial X_{mn,j}}{\partial t} =  F^{mn}_j
\end{equation}

where $\tau$ is chosen to be on the timescale of the least damped eigenmode. VMEC, in fixed-boundary mode, then takes as inputs the pressure and either the rotational transform or the net toroidal current profile as flux functions (The rotational transform is given by $\iota(s) = \chi'/\psi'$), along with the Fourier series describing the desired boundary shape, $R_b(u,v), Z_b(u,v)$.

\subsection{DESC}\label{DESC}

DESC\citep{dudt_desc_2020}, another 3D equilibrium code developed recently, is a pseudospectral code that finds equilibria by minimizing the MHD force balance error \eqref{eq:F} directly at collocation nodes, as opposed to minimizing energy through a variational principle. Similar to VMEC, the base geometry is a cylindrical coordinate system $\mathbf{x} = (R,\phi,Z)$. DESC uses as its computational grid the coordinates $\bm{\alpha}_{DESC} = (\rho,\theta,\zeta)$, with $\rho$ being a radial coordinate proportional to the square root of the normalized toroidal flux, $\theta$ a poloidal angle, and $\zeta$ is the geometric toroidal angle, the same angles as are used by VMEC (note that this is different than the original publication \citep{dudt_desc_2020}, which used the straight-field-line $\theta^*$ in the computational domain):

\begin{subequations}
\begin{equation}
    \rho = \sqrt{\frac{\psi}{\psi_a}} \hspace{2mm}, \hspace{2mm} 0\leq \rho \leq 1
\end{equation}
\begin{equation}
    \theta = \theta^* - \lambda(\rho,\theta,\zeta)\hspace{2mm}, \hspace{2mm} 0\leq \theta \leq 2\pi
\end{equation}
\begin{equation}
    \zeta = \phi  \hspace{2mm}, \hspace{2mm} 0\leq \zeta \leq 2\pi/N_{FP}
\end{equation}
\end{subequations}

where, similar to VMEC, $\lambda(\rho,\theta,\zeta)$ is a function periodic in $(\theta,\zeta)$ that converts $\theta$ to a magnetic poloidal angle $\theta^*$ \citep{helander_theory_2014}.

DESC, like VMEC, solves the inverse equilibrium problem. Unlike VMEC, DESC expands $R(\rho,\theta,\zeta),Z(\rho,\theta,\zeta),\lambda(\rho,\theta,\zeta)$ in spectral bases in all three coordinates, using a Fourier series toroidally and Zernike polynomials in the radial and poloidal directions \citep{zernike_diffraction_1934, sakai_application_2009}:

\begin{subequations}\label{eq:zernike_basis}
\begin{equation}
    R(\rho,\theta,\zeta) = \sum_{m=-M}^M \sum_{n=-N}^N \sum_{l=0}^{L} R_{lmn} \mathcal{Z}_l^m (\rho,\theta) \mathcal{F}^n(\zeta)
\end{equation}
\begin{equation}
    \lambda(\rho,\theta,\zeta) = \sum_{m=-M}^M \sum_{n=-N}^N \sum_{l=0}^{L} \lambda_{lmn} \mathcal{Z}_l^m (\rho,\theta) \mathcal{F}^n(\zeta)
\end{equation}
\begin{equation}
Z(\rho,\theta,\zeta) = \sum_{m=-M}^M \sum_{n=-N}^N \sum_{l=0}^{L} Z_{lmn} \mathcal{Z}_l^m (\rho,\theta) \mathcal{F}^n(\zeta)
\end{equation}
\end{subequations}

Where $\mathcal{Z}_l^m$ is the Zernike polynomial of radial degree $l$ and poloidal degree $m$, defined as:

\begin{equation}
    \mathcal{Z}_m^l(\rho,\theta) = \begin{cases} \mathcal{R}_l^{|m|}(\rho) cos(|m|\theta) & \text{for } m\geq 0 \\
    \mathcal{R}_l^{|m|}(\rho) sin(|m|\theta) & \text{for } m < 0 
    
    \end{cases}
\end{equation}

With the radial function $\mathcal{R}_l^{|m|}$ as the shifted Jacobi polynomial:

\begin{equation}
\mathcal{R}_l^{|m|} (\rho) = \sum_{s=0}^{(l-|m|)/2} \frac{(-1)^s(l-s)!}{ s!\left[ (l+|m|)/2 - s\right]! \left[ (l-|m|)/2 + s\right]!  }\rho^{l-2s}
\end{equation} 

And $\mathcal{F}$ is the typical Fourier series in $\zeta$:

\begin{equation}
    \mathcal{F}^n(\zeta) = \begin{cases} cos(|n|N_{FP}\zeta) \text{for } n\geq 0 \\
    sin(|n|N_{FP}\zeta) \text{for } n < 0 
    \end{cases}
\end{equation}

The basis vector and Jacobian definitions given in Eqs. \eqref{eq:switch_and_duality} and \eqref{eq:jacobian} have obvious analogues with the DESC coordinate system, with $(s \rightarrow \rho, u \rightarrow \theta, v \rightarrow \zeta)$. It is worth noting that the choice of Zernike polynomials in the spectral basis ensures analyticity at the magnetic axis. Any analytic function when expanded in a Fourier series near the origin of a disk must have a radial structure that goes as \citep{lewis_physical_1990}:
\begin{equation}
    a_m(\rho) = \rho^m (a_{m,0} + a_{m,2}\rho^2 + a_{m,4}\rho^4 +...)
\end{equation}

where $a_{m,i}$ is the $i$th term in a Taylor series expansion of the $m$th poloidal Fourier coefficient $a_m(\rho)$. With the Zernike basis, any spectral coefficient with poloidal mode number $m$ necessarily has a radial dependence that scales as $\rho^m$, thus inherently satisfying this constraint and ensuring only physical modes are included in the spectrum of $R$ and $Z$. DESC employs the same nested flux surfaces assumption as VMEC to arrive at a similar contravariant form of the magnetic field:

\begin{align}
    \bB &= \nabla \rho \times \nabla \theta^* + \nabla \zeta \times \nabla \chi\\
    &= B^{\theta} \mathbf e_{\theta} + B^{\zeta} \eze
\end{align}

With contravariant components given by:

\begin{subequations}
\begin{align}
    B^\rho &= 0, ~~\text{due to}~\bB\cdot \eR = 0\\
   B^\theta &= \frac{1}{\sqrt{g}} \Big(\chi' - \psi' \lze \Big)\\
B^\zeta &= \frac{1}{\sqrt{g}} \psi'\Big(1 + \lth \Big)
\end{align}
\end{subequations}

The MHD force balance equation is:

\begin{equation}
    \bm{F} = F_{\rho} \nabla \rho + F_{\beta} \bm{\beta_{DESC}}
\end{equation}

with the two independent force components:

\begin{subequations}
\begin{align}
    F_{\rho} &= \sqrt{g} (J^{\zeta} B^{\theta} - J^{\theta} B^{\zeta}) + p'\\
    F_{\beta} &= \sg J^{\rho}
\end{align}
\end{subequations}

and the vector $\bm{\beta_{DESC}}$ in the helical direction:

\begin{equation}
    \bm{\beta_{DESC}} = B^{\zeta} \nabla \theta - B^{\theta} \nabla \zeta
\end{equation}

which is the same direction as the VMEC $\bm{\beta}$, but without the factor of $\sg$. The current density components are found with Eq. \eqref{eq:J_VMEC}, with DESC coordinates $\alpha_i = (\rho,\theta,\zeta)$. By weighting the force components by volume, one can obtain a system of equations for the total MHD force balance error in the plasma volume \citep{dudt_desc_2020}:

\begin{subequations}\label{eq:f_weighted_DESC}
\begin{equation}
    f_{\rho} = F_{\rho} ||\nabla \rho||_2 \sg \Delta \rho \Delta \theta \Delta \zeta
\end{equation}
\begin{equation}
    f_{\beta} = F_{\beta} ||\bm{\beta}||_2 \sg \Delta \rho \Delta \theta \Delta \zeta
\end{equation}
\end{subequations}

As a pseudospectral code, DESC solves for the equilibrium by solving the force balance error equations \eqref{eq:f_weighted_DESC}, evaluated on a collocation grid. DESC solves the resulting nonlinear system of equations $\bm{f}(\bm{x}) = [f_{\rho},f_{\beta}](\bm{x})=\bm{0}$, where $\bm{x} = [R_{lmn},Z_{lmn},\lambda_{lmn}]$ are the coefficients of the spectral representation (given in Eq. \eqref{eq:zernike_basis}) of the flux surface positions and the stream function $\lambda$. Newton-Raphson type methods from Scipy \citep{virtanen_scipy_2020} such as Levenberg-Marquadt are employed as the nonlinear equation solver in DESC, which can achieve quadratic convergence near the solution \citep{press_numerical_1996}. It is worth noting that DESC is flexible enough to find equilibria by minimizing different objective functions, such as MHD energy, but it has been found in extensive testing that using force error is faster (by a factor of two in some cases) and yields better convergence in solving for MHD equilibria in DESC, and so force error minimization is used for the DESC results in this paper. Force is believed to be a better objective for solving equilibria with DESC's pseudospectral formulation because it allows the code to take advantage of local information afforded by the force balance equation, which is already evaluated on the collocation nodes due to the pseudospectral approach. Finally, DESC is similar to VMEC in that, in fixed-boundary mode, it takes as inputs the pressure profile and either the rotational transform or the net toroidal current profile as flux functions (The rotational transform is given by $\iota(\rho) = \chi'/\psi'$), along with the Fourier series describing the desired boundary shape, $R_b(\theta,\zeta), Z_b(\theta,\zeta)$.\\

\section{Comparison Methods}\label{methods}

In order to compare the two equilibrium codes, a common metric must be used. VMEC explicitly minimizes MHD energy using a gradient descent method, while DESC minimizes MHD force error in the plasma volume. To compare the two code results, the resulting solution MHD force balance error will be shown, as well as time-to-solution. To verify the correlation of low force error with accurate calculations of interesting physics metrics, Mercier stability  calculated with both codes and compared to asymptotic near-axis formulas \citep{landreman_magnetic_2020} will also be shown in section \ref{mercier_section}.\\
VMEC does not output the force balance error in real space, so it was calculated from the VMEC-outputted Fourier coefficients of $R,Z,\lambda$. The derivation of the equations used for VMEC force balance is given in Appendix \ref{appendix:F_bal_deriv_VMEC}. Once the force error at each point in $(s,u,v)$ space was calculated, both volumetric and flux-surface averages were calculated. The volume average was calculated as:

\begin{equation}\label{eq:F_vol_avg}
    \langle F \rangle_{vol} = \frac{\int_{\theta=0}^{2\pi}\int_{\phi=0}^{2\pi}\int_{s=0.1}^{0.99} |F||\sqrt{g}|ds d\phi d\theta }{\int_{\theta=0}^{2\pi}\int_{\phi=0}^{2\pi}\int_{s=0.1}^{0.99} |\sqrt{g}|ds d\phi d\theta }
\end{equation}

Where the radial integration does not include the axis or edge to avoid sensitivities of the force error calculation at these locations. It should be noted that DESC solutions did not have this limitation, and are integrated throughout the entire volume, while the VMEC solutions are only integrated through the above range. This difference could make the VMEC calculations appear to have lower error than the full volume integration would otherwise yield. The flux surface average at a given radial position $s$ was calculated as:
\begin{equation}
    \langle F\rangle_{fsa}(s) = \frac{\int_{\theta=0}^{2\pi}\int_{\phi=0}^{2\pi} |F(s)||\sqrt{g}(s)| d\phi d\theta }{\int_{\theta=0}^{2\pi}\int_{\phi=0}^{2\pi} |\sqrt{g}(s)|d\phi d\theta }
\end{equation}

Then to yield a normalized, unitless error metric, the above quantities are divided by the volume average of the pressure gradient magnitude:

\begin{subequations}
\begin{equation}
    |\nabla p| = \sqrt{\left(\frac{dp}{ds} \right)^2 \bm{e}^s \cdot \bm{e}^s} = \sqrt{\left(\frac{dp}{ds} \right)^2 g^{ss}}
\end{equation}
\begin{equation}
    \langle |\nabla p|\rangle_{vol}=\frac{\int_{\theta=0}^{2\pi}\int_{\phi=0}^{2\pi}\int_{s=0}^{1} |\nabla p||\sqrt{g}|ds d\phi d\theta }{\int_{\theta=0}^{2\pi}\int_{\phi=0}^{2\pi}\int_{s=0}^{1} |\sqrt{g}|ds d\phi d\theta }
\end{equation}

\end{subequations}

\begin{figure}
    \centering
    \includegraphics[keepaspectratio, width=3in]{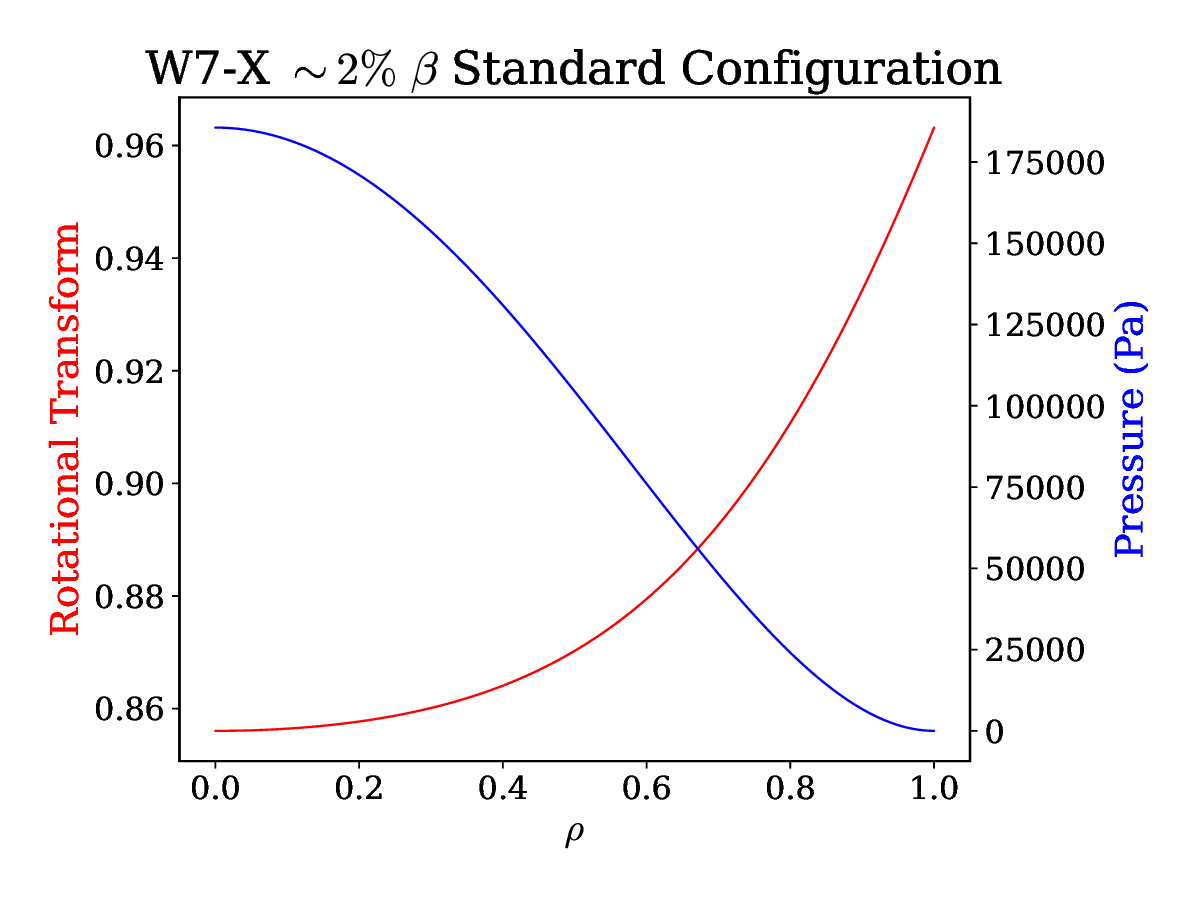}
    \caption{Pressure and rotational transform profiles used as inputs for the fixed-boundary W7-X standard configuration equilibria computed in this paper.}
    \label{fig:W7X_profiles}
\end{figure}

With the above normalized error metrics defined, both codes were ran in fixed-boundary, fixed-iota mode to solve equilibria for a W7-X standard configuration, finite beta ($\beta \approx 2\%$) equilibrium \citep{sunn_pedersen_plans_2015}. The equilibrium used in this paper had the following rotational transform and pressure profiles (given as a power series in $\rho = \sqrt{s} = \sqrt{\frac{\psi}{\psi_a}}$):
\begin{align}
    p(\rho) &= 185596.929 - 371193.859 \rho^2 \\
    &+ 185596.929 \rho^4 ~ [Pa]\\
    \iotabar(\rho) &= 0.856047021 + 0.0388095412 \rho^2 + 0.0686795128 \rho^4 \\
    &~~+ 0.0186970315 \rho^6 - 0.0190561179 \rho^8\\
\end{align}
These profiles are plotted in Figure \ref{fig:W7X_profiles}. The full base input files for VMEC and DESC that the runs in this paper are based off of are available in the DESC Github repository \citep{Dudt_DESC}, which include the boundary shape Fourier series, which goes up to $M=N=12$.


\subsection{VMEC Radial Derivative}

In VMEC, the outputs $(R_{mn},Z_{mn},L_{mn})$, from which derived quantities of magnetic field, current density and ultimately force balance error can be calculated, are given on a discrete radial grid. To calculate the force balance error Eq. \eqref{eq:F}, derivatives of $R,Z$ up to second order in each of $(s,u,v)$ are necessary, and in the radial direction these derivatives must be found numerically. A comparison of different numerical derivative methods was carried out, to see the sensitivity of the resulting force balance error to the method used. The radial derivatives were carried out on the Fourier coefficients $(R_{mn},Z_{mn},L_{mn})$. \\

\begin{figure}
    \centering
    \includegraphics[keepaspectratio,width=3.5in]{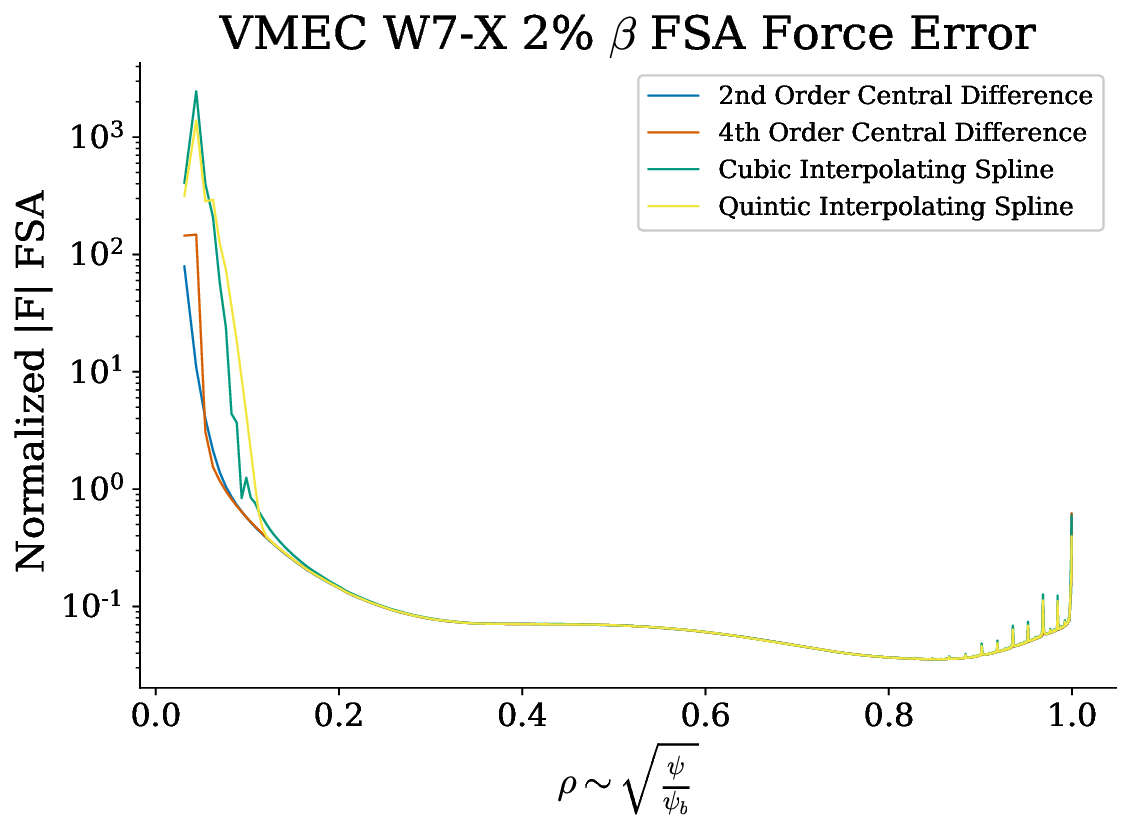}
    \caption{W7-X flux surface average of normalized force error versus $\rho$ with different radial derivative methods. All have angular resolution of M=N=16 and NS=1024 flux surfaces.}
    \label{fig:compare_radial_derivs_FSA}
\end{figure}

Figure \ref{fig:compare_radial_derivs_FSA} shows the normalized flux-surface averaged force balance error calculated for a VMEC W7-X equilibrium with $M=N=16$ angular resolution and $ns=1024$ flux surfaces using several different numerical derivative methods: finite difference (2nd-order and 4th-order central differences \citep{collatz_numerical_1960}), and cubic and quintic interpolating splines. It can be seen that the numerical method used does not impact the calculated force error in the majority of the plasma, and mainly changes the calculated force error near the magnetic axis. As there is such a large sensitivity in the force error at axis to the numerical method used, for all volume averages of force error from VMEC, the radial integration is limited to $s = 0.1 \rightarrow 0.99$, in order to avoid including this sensitive portion of the calculation.\\

\begin{figure}
    \centering
    \includegraphics[keepaspectratio, width=4.25in]{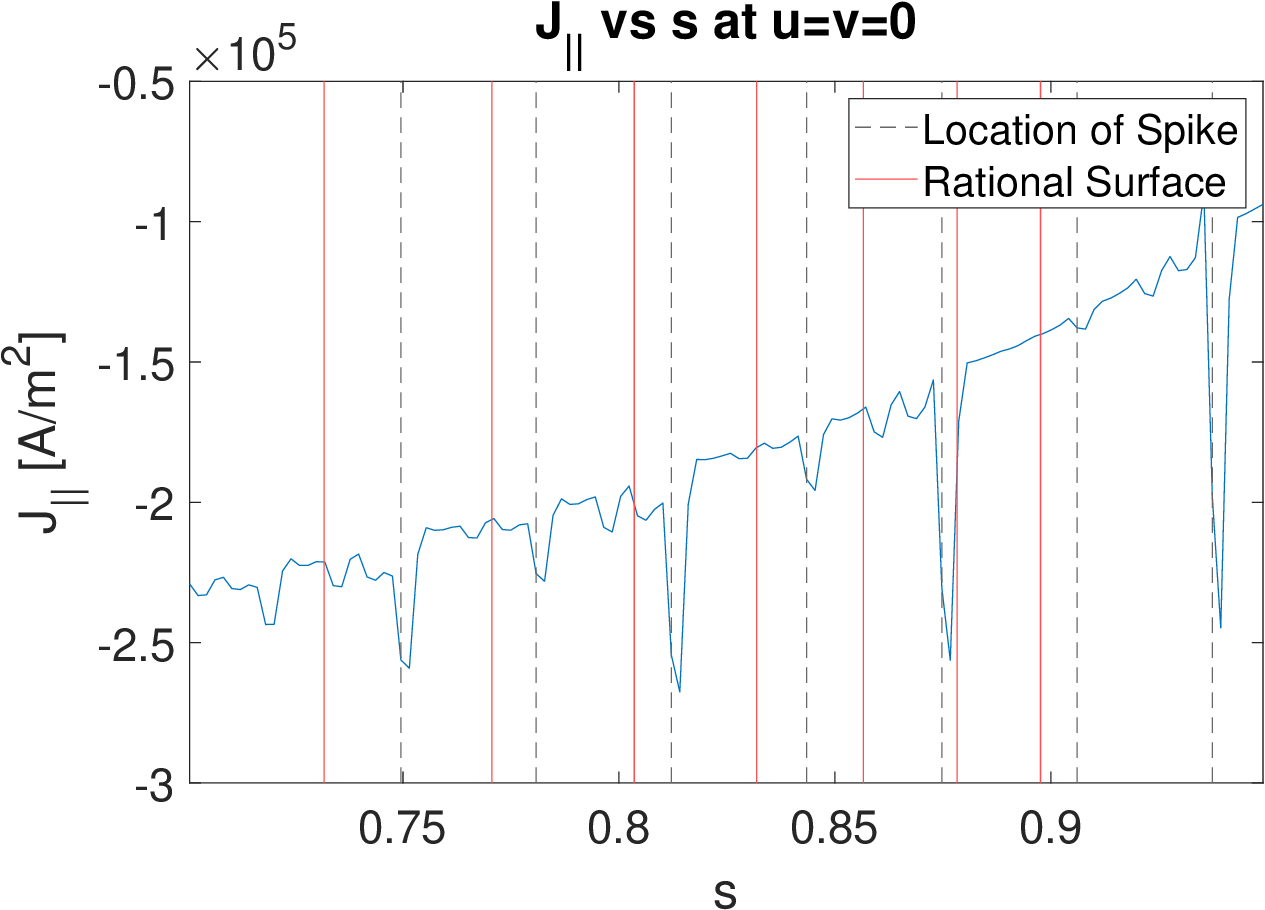}
    \caption{Parallel current density plotted versus normalized toroidal flux $s$ at $u=v=0$ for a W7-X-like equilibrium solved in VMEC with M=N=16 and ns=512. Also shown are the low order rational surface locations, as well as the locations of the spikes in the force error shown in Figure \ref{fig:compare_radial_derivs_FSA}}
    \label{fig:Jparallel_spikes}
\end{figure}

Additionally, there are noticeable spikes observed in the calculated force error of these solutions near the edge, which stem from spikes in the VMEC current densities at those locations. These spikes were observed to appear at locations in $s$ corresponding to coarser grids used in the continuation method (i.e. NS\_ARRAY=[16, 32, 64, 128, 256, 512, 1024], and the spikes are at locations corresponding to the NS=32 grid). They appear as discontinuous jumps and spikes in the first and second radial derivative of the $R,Z$ Fourier coefficients, which propagate to the current density and force error. It is speculated that these are due to convergence issues with the highly shaped equilibrium. These are not due to issues at rational surfaces, as Figure \ref{fig:Jparallel_spikes} plots the parallel current density versus $s$ along $u=v=0$, along with low-order rationals. This figure shows that the spikes do not line up with the rational surfaces, and so are not due to the rational surfaces. Shown in Figure \ref{fig:ftol_scan} in Appendix \ref{appendix:spikes} are results of running VMEC with increased solver tolerance, and Figure \ref{fig:F_fsa_MN_conv} shows VMEC runs with higher angular resolution, neither of which completely alleviate the issue. However, for the purposes of this comparison, the spikes are localized enough that they do not significantly affect the volume-averaged error.

\section{Results}\label{results}

\subsection{Spectral Properties}

\begin{figure}
    \centering
    \includegraphics[keepaspectratio,width=5in]{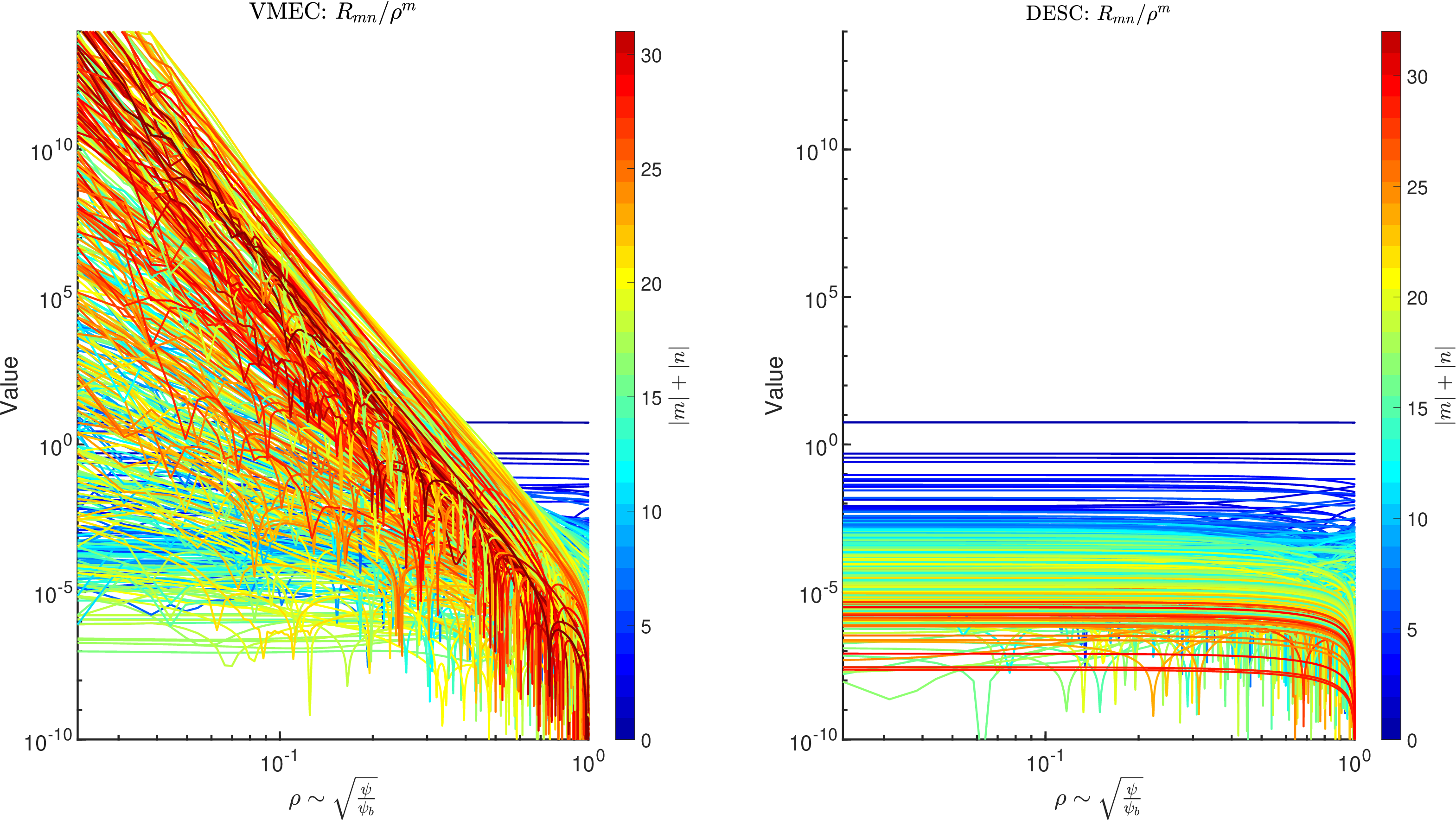}
    \caption{Physical constraint on Fourier coefficients near axis, where an analytic function's Fourier series coefficients should scale as $\rho^m$ as they approach the origin \citep{lewis_physical_1990}. Note the diverging of the VMEC coefficients when divided by $\rho^m$ as the axis is approached, indicating that they do not satisfy this analyticity constraint, leading to unphysical modes near axis.}
    \label{fig:analytic_rhom_origin}
\end{figure}

\begin{figure}
    \centering
    \includegraphics[keepaspectratio,width=3in]{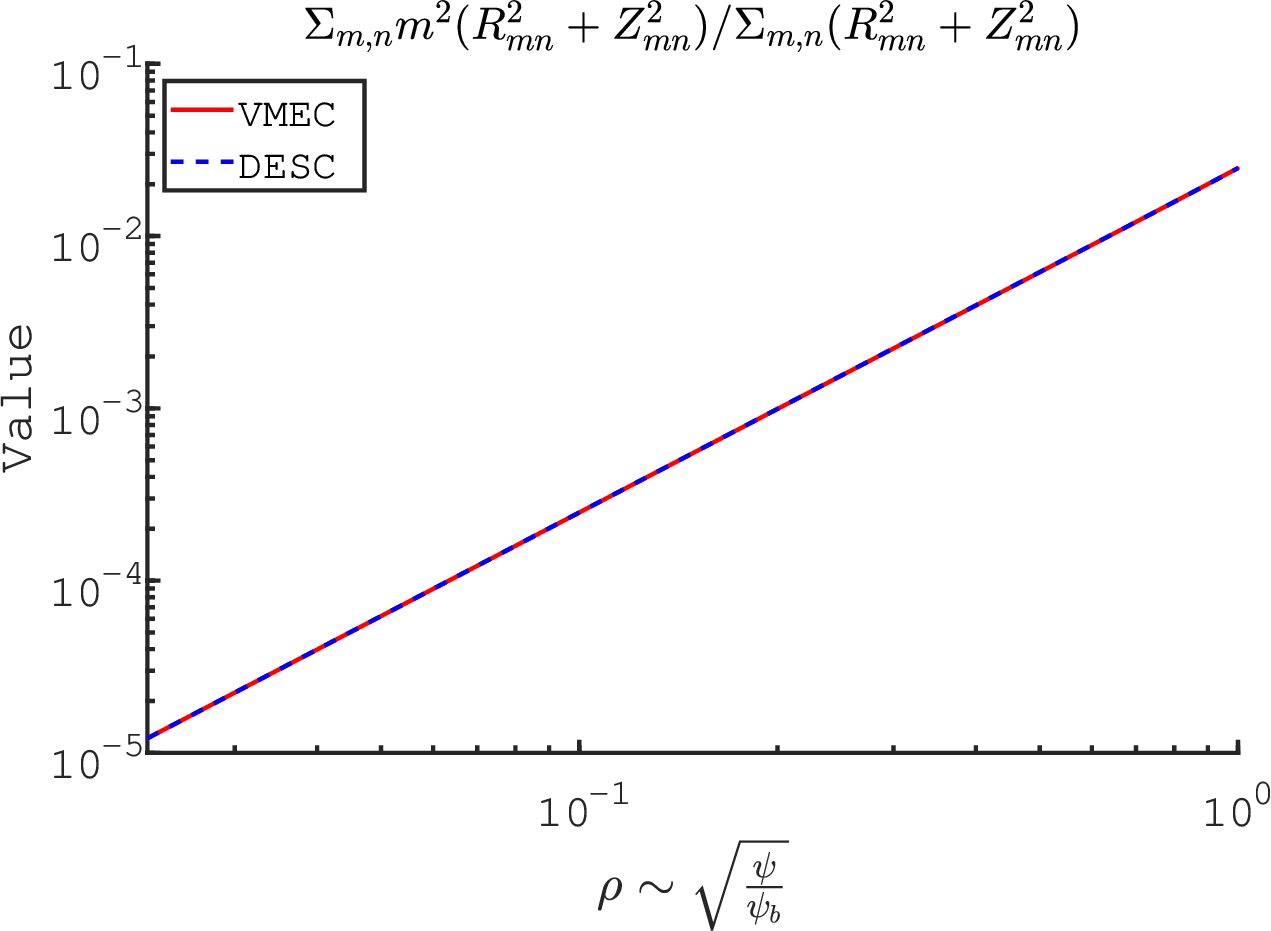}
    \caption{Spectral width ($p=q=2$) of the VMEC and DESC spectra for a W7-X equilibrium. It can be seen that DESC, while not explicitly enforcing any poloidal angle constraints, ends up finding an optimal representation through the course of the optimization procedure. The equilibrium solved is the W7-X standard configuration at $\beta=2\%$  with $M=N=16$ angular resolution, $ns=2048$ for the VMEC solution and $L=16$ for the DESC solution}
    \label{fig:spec_width}
\end{figure}

To compare the spectral representations of the two codes, the radial dependence of the spectral coefficients of $R$ and the spectral width defined in Eq. \eqref{SpecWidth} were calculated and compared. Figure \ref{fig:analytic_rhom_origin} shows the amplitude of each $R_{mn}$ Fourier coefficient factored by its $\rho^m$ dependence (the DESC solution was transformed here from a global Fourier-Zernike to a Fourier basis on discrete flux surfaces to compare directly to the VMEC solution). The DESC coordinate $\rho = \sqrt{\frac{\psi}{\psi_a}}$ was factored out from both codes' coefficients because this radial variable is proportional to the typical polar radius $r$. It can be seen from the figure that while the DESC Fourier coefficient amplitudes are relatively constant with $\rho$, indicative of the correct scaling necessary for analyticity at the origin, the VMEC higher order mode amplitudes tend to diverge near $\rho=0$. This is evidence of possibly unphysical modes existing near-axis in the VMEC Fourier spectrum.
\\
As a further point of comparison, Figure \ref{fig:spec_width} shows the spectral width metric calculated for a DESC and a VMEC W7-X-like finite beta solution. The spectral width is essentially the same for each code, which is indicative of the stream function $\lambda$ being chosen so as to optimize the Fourier spectrum representation of the flux surfaces. VMEC does this through the Hirshman-Breslau constraint \citep{hirshman_explicit_1998}, while DESC lets $\lambda$ vary through the course of solving, and the optimization routines arrive at an optimal $\lambda$. 

\subsection{Numerical Convergence}

To compare the convergence of the VMEC and DESC codes with respect to radial resolution, a convergence study with each code was carried out using an axisymmetric D-shaped equilibrium similar to that in \citet{hirshman_steepestdescent_1983} and \citet{dudt_desc_2020}. The equilibrium used in this paper had the following rotational transform and pressure profiles (given as a power series in $\rho = \sqrt{s} = \sqrt{\frac{\psi}{\psi_a}}$):

\begin{align}
    p(\rho) &= 1600 - 3200 \rho^2 + 1600 \rho^4 ~ [Pa]\\
    \iotabar(\rho) &= 1 - 0.67 \rho^2
\end{align}

\begin{figure}
    \centering
    \includegraphics[keepaspectratio, width=3.5in]{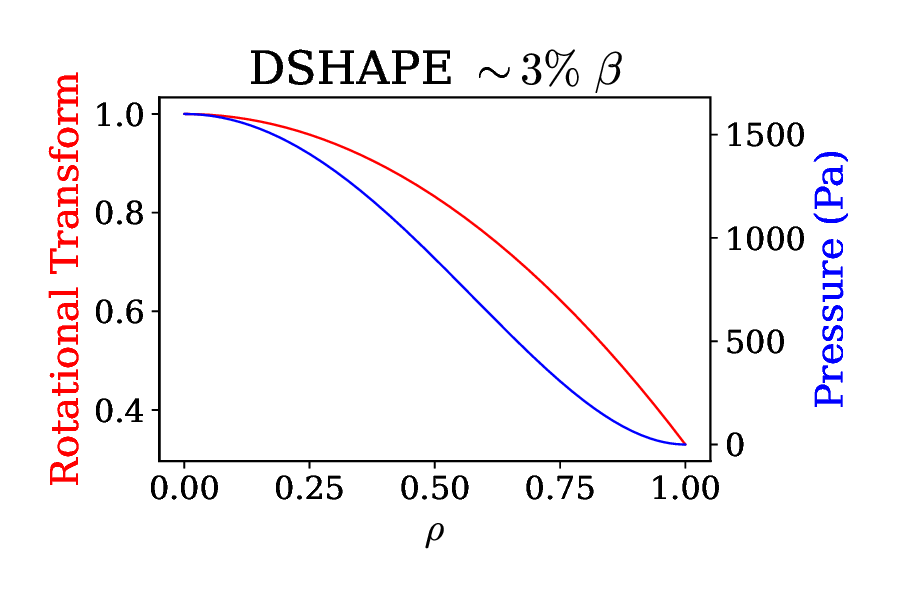}
    \caption{Pressure and rotational transform profiles used as inputs for the fixed-boundary D-shaped equilibria computed in this paper.}
    \label{fig:DSHAPE_profiles}
\end{figure}

These profiles are plotted in Figure \ref{fig:DSHAPE_profiles}. The boundary shape and enclosed flux (in Wb) are given by:
\begin{align*}
R^b &= 3.51 - \cos\theta + 0.106 \cos2\theta \\
Z^b &= 1.47 \sin\theta + 0.16 \sin2\theta \\
\psi_a &= 1
\end{align*}
The full D-shaped input files for VMEC and DESC that the runs in this section are based off of are available in the DESC Github repository \citep{Dudt_DESC}.


\begin{figure}
    \centering
    \includegraphics[keepaspectratio,width=3.in]{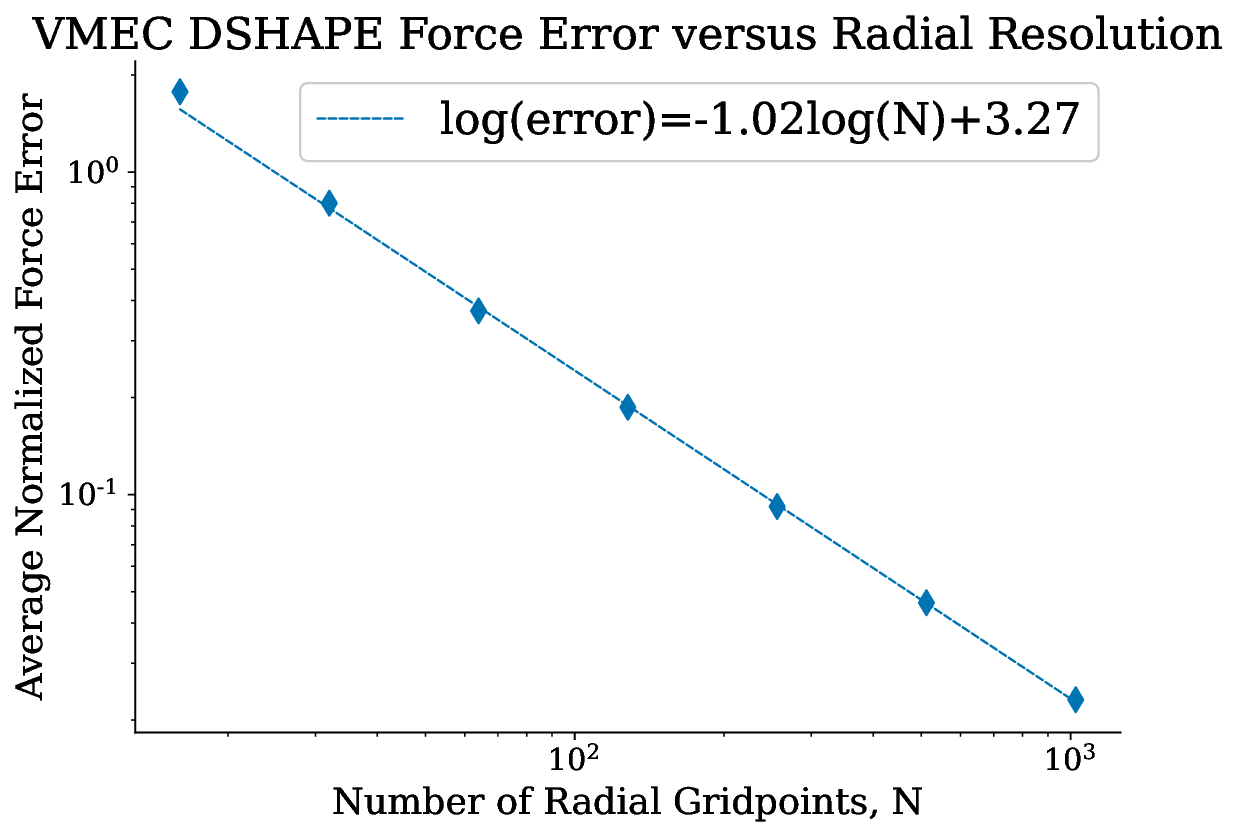}
    \caption{D-shaped M=16 error convergence with increasing radial resolution in VMEC, on a log-log scale. Note the first order convergence rate, due to the first order finite differences used in the radial direction.}
    \label{fig:VMEC_DSHAPE_Radial_conv}
\end{figure}

\begin{figure}
    \centering
    \includegraphics[keepaspectratio,width=3in]{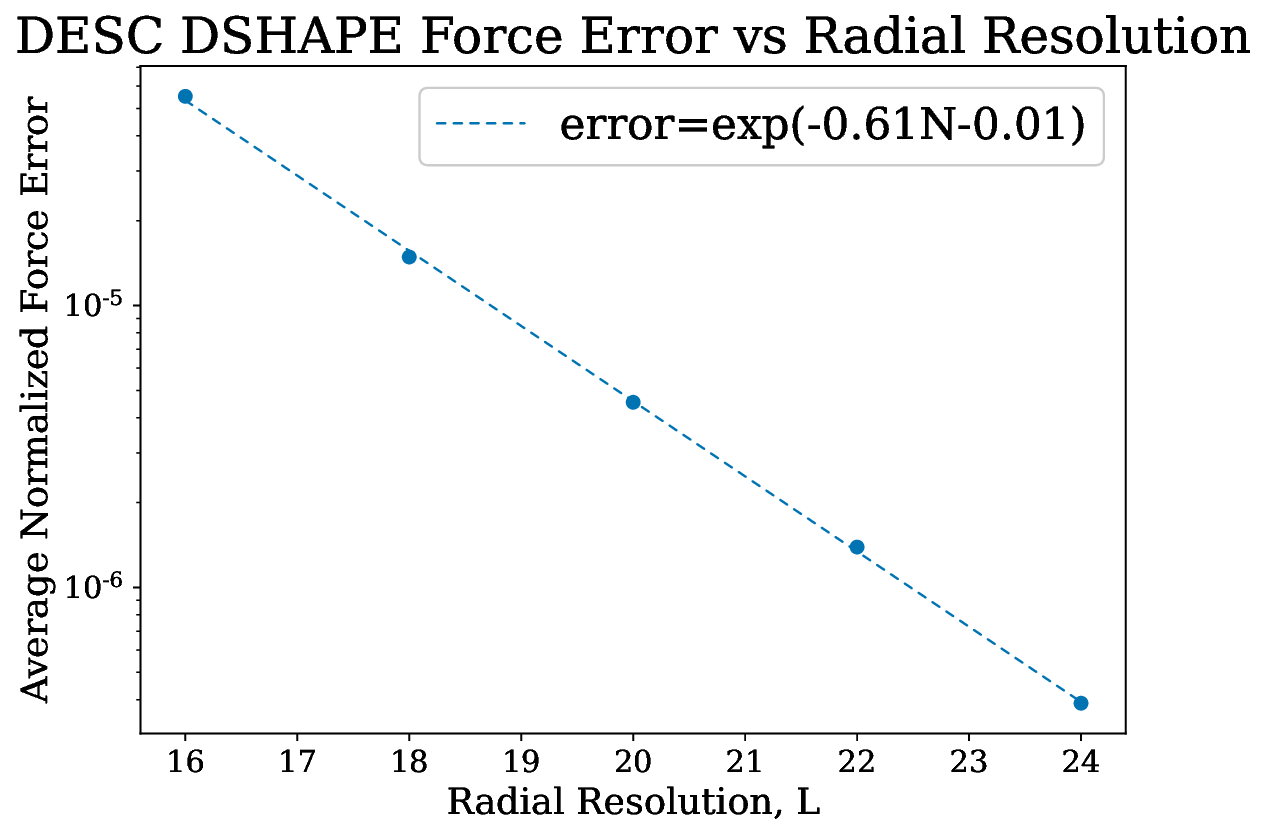}
    \caption{D-shaped M=16 error convergence for increasing radial resolution in DESC, on a semi-log scale. Note that the linearity here is indicative of exponential convergence.}
    \label{fig:DESC_DSHAPE_Radial_conv}
\end{figure}

\begin{figure}
    \centering
    \includegraphics[keepaspectratio, width=3.5in]{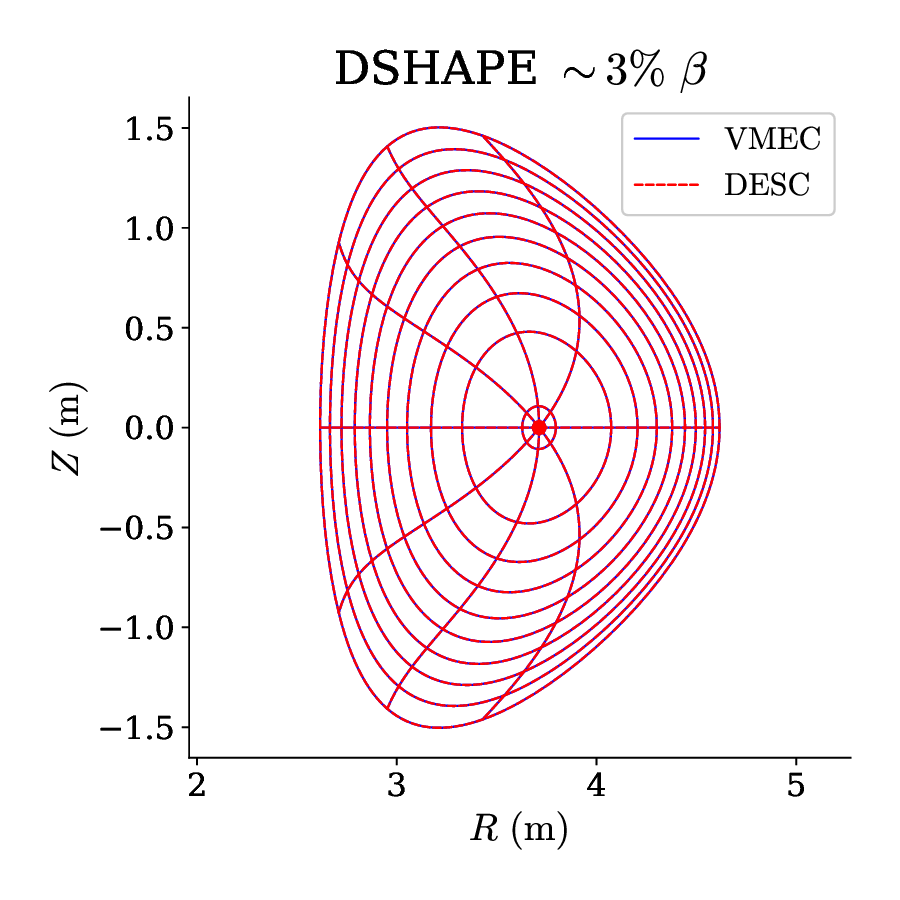}
    \caption{Flux surfaces for a VMEC (ns=1024 M=16) and a DESC (L=M=16) D-shaped equilibrium solution.}
    \label{fig:DSHAPE_surf_comp}
\end{figure}

Both the VMEC and the DESC codes use a spectral representation for the angular dependence of their solutions. Consequently, we would expect, assuming a smooth solution, that the error convergence will be exponential with increasing angular resolution \citep{boyd_chebyshev_2001}. However, the codes differ in the radial direction, as VMEC's solution is explicitly represented only on a finite grid and the code employs a first-order finite difference scheme, while DESC's spectral representation describes the solution radially as well as in angle. Thus, we would expect that the radial finite differences in VMEC would limit the radial convergence to be first order, while in DESC we should still see exponential convergence with increasing radial resolution (again, given a smooth solution). This is summarized in table \ref{tab:DESC_VMEC_Convergence}.\\
 Figure \ref{fig:DSHAPE_surf_comp} shows the agreement of the flux surfaces between the VMEC solution with $ns=1024$ and the DESC solution with $L=16$, indicating that both codes resolve the same solution qualitatively. Figure \ref{fig:VMEC_DSHAPE_Radial_conv} shows the average normalized force error of a D-shaped solution found with VMEC versus increasing radial resolution. The poloidal resolution for these runs was kept fixed at $M=16$ in both codes. Plotted on a log-log scale, the best-fit line's slope of $-1.02$ clearly shows the first-order convergence of VMEC with increasing radial resolution, as expected. 
Figure \ref{fig:DESC_DSHAPE_Radial_conv} shows the average normalized force error in a DESC D-shaped solution for increasing radial spectral resolution $L$, for fixed poloidal resolution $M=16$. Note that here the plot is now a \emph{log-linear} scale, where the linear dependence of error on resolution indicates that the error convergence of the DESC solution is \emph{exponential} with increasing radial resolution, as shown in the legend of the exponential fit of the error to the radial resolution. Through its Fourier-Zernike spectral basis, the DESC code is able to achieve exponential convergence with increasing radial resolution, a scaling unattainable with the limitations imposed by the radial first-order finite differences in the VMEC code.


\begin{table}
    \centering
    \begin{tabular}{ccc}
         & \textbf{Angular Convergence} & \textbf{Radial Convergence}   \\\hline
    \textbf{DESC}     &  Exponential & \textcolor{ForestGreen}{Exponential} \\\hline
    \textbf{VMEC} & Exponential & \textcolor{red}{Algebraic $\bm{O}(N_{radial}^{-1})$} \\ \hline
    \end{tabular}
    \caption{Expected convergence with respect to each resolution parameter for VMEC and DESC}
    \label{tab:DESC_VMEC_Convergence}
\end{table}



\subsection{W7-X Solution Comparison}

\subsubsection{Flux Surface Comparison}
\begin{figure*}
    \centering
    \includegraphics[keepaspectratio, width=\columnwidth
    ]{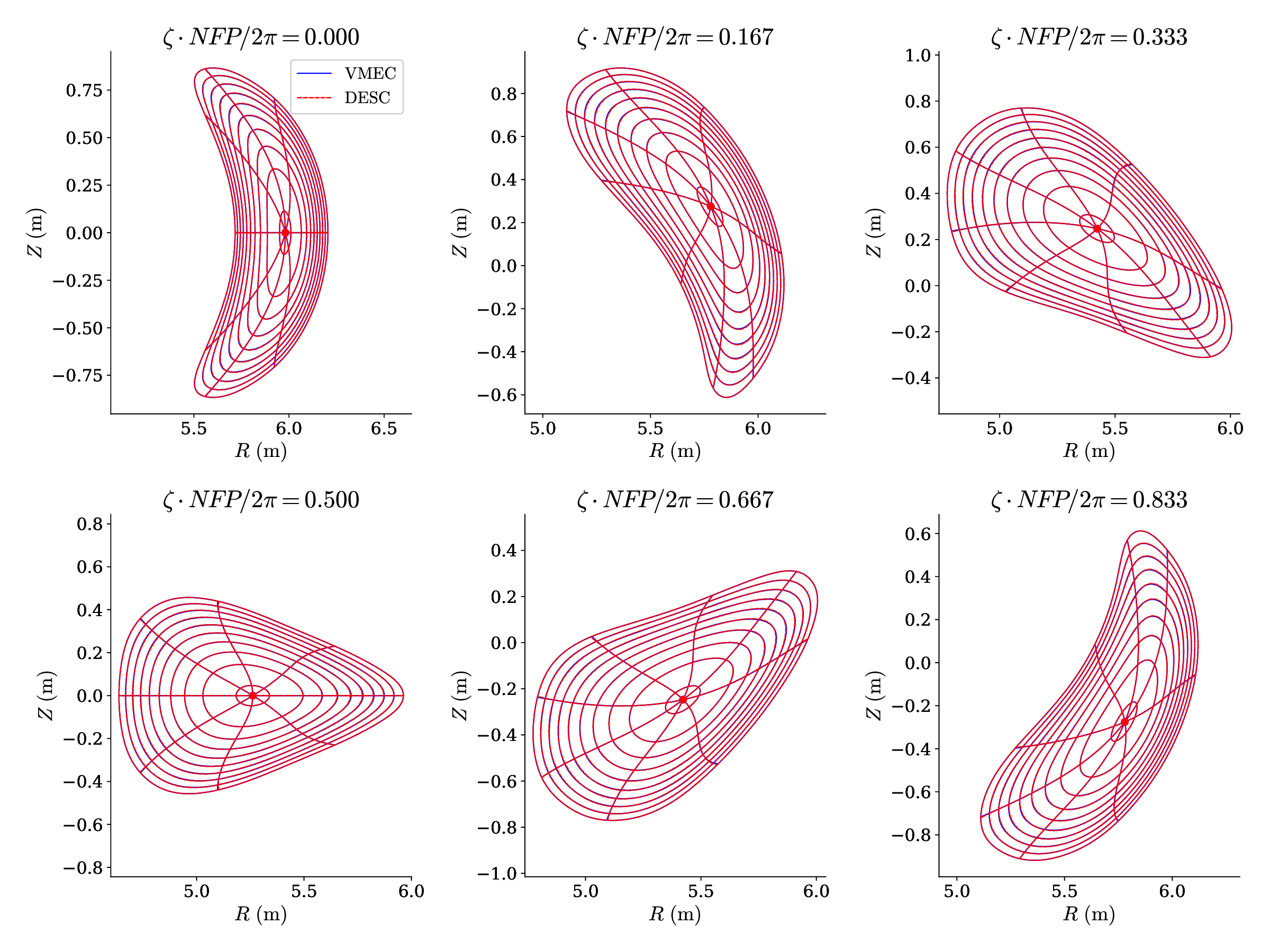}
    \caption{Flux surfaces for a VMEC (ns=1024 M=N=16) and a DESC (L=M=N=16) W7-X equilibrium solution.}
    \label{fig:W7X_surfaces}
\end{figure*}

The flux surfaces of the W7-X equilibrium solution found by DESC ($L=M=N=16$, in red) and VMEC ($ns=1024$ $M=N=16$, in blue) are shown in Figure \ref{fig:W7X_surfaces}. The overlap of the surfaces shows the agreement of the two codes, indicating that they resolve qualitatively similar solutions to the equilibrium problem.

\subsubsection{Force Error}

\begin{figure}
    \centering
    \includegraphics[keepaspectratio, width=3.6in]{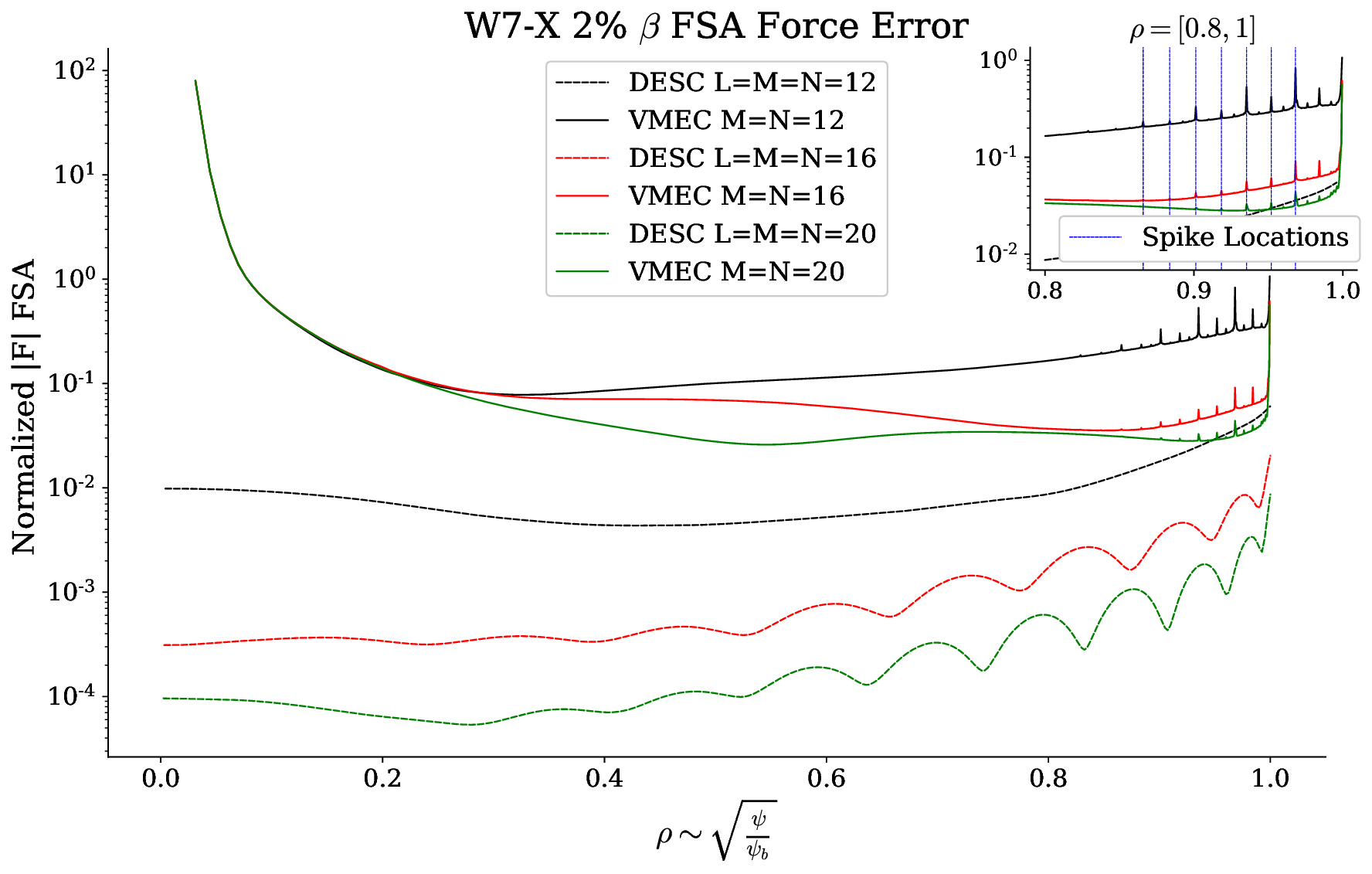}
    \caption{W7-X Flux surface average of normalized force error versus $\rho$ for increasing VMEC angular resolution (all with radial resolution of 1024 flux surfaces) along with DESC solution. 2nd order finite differences were used for the radial derivatives in the VMEC force error calculation. The insert shows that the error spikes occur at the same radial position for each VMEC solution shown, independent of resolution.}

    \label{fig:F_fsa_VMEC_DESC}
\end{figure}

The normalized force error flux surface average is compared directly between DESC and VMEC in Figure \ref{fig:F_fsa_VMEC_DESC}. Here, the VMEC solutions noticeably have their largest force error as they approach the axis, while the DESC solutions maintain accuracy in satisfying force balance near-axis. This inaccuracy of the VMEC solutions near axis could be attributed to the modes in its Fourier spectrum which lack the correct radial scaling (shown earlier in Figure \ref{fig:analytic_rhom_origin}). 
The oscillations seen in the higher resolution DESC solutions correspond to the collocation points -- lower force balance errors are expected on the surfaces where the residuals were minimized. 

\begin{figure}
    \centering
    \includegraphics[keepaspectratio, width=3.25in]{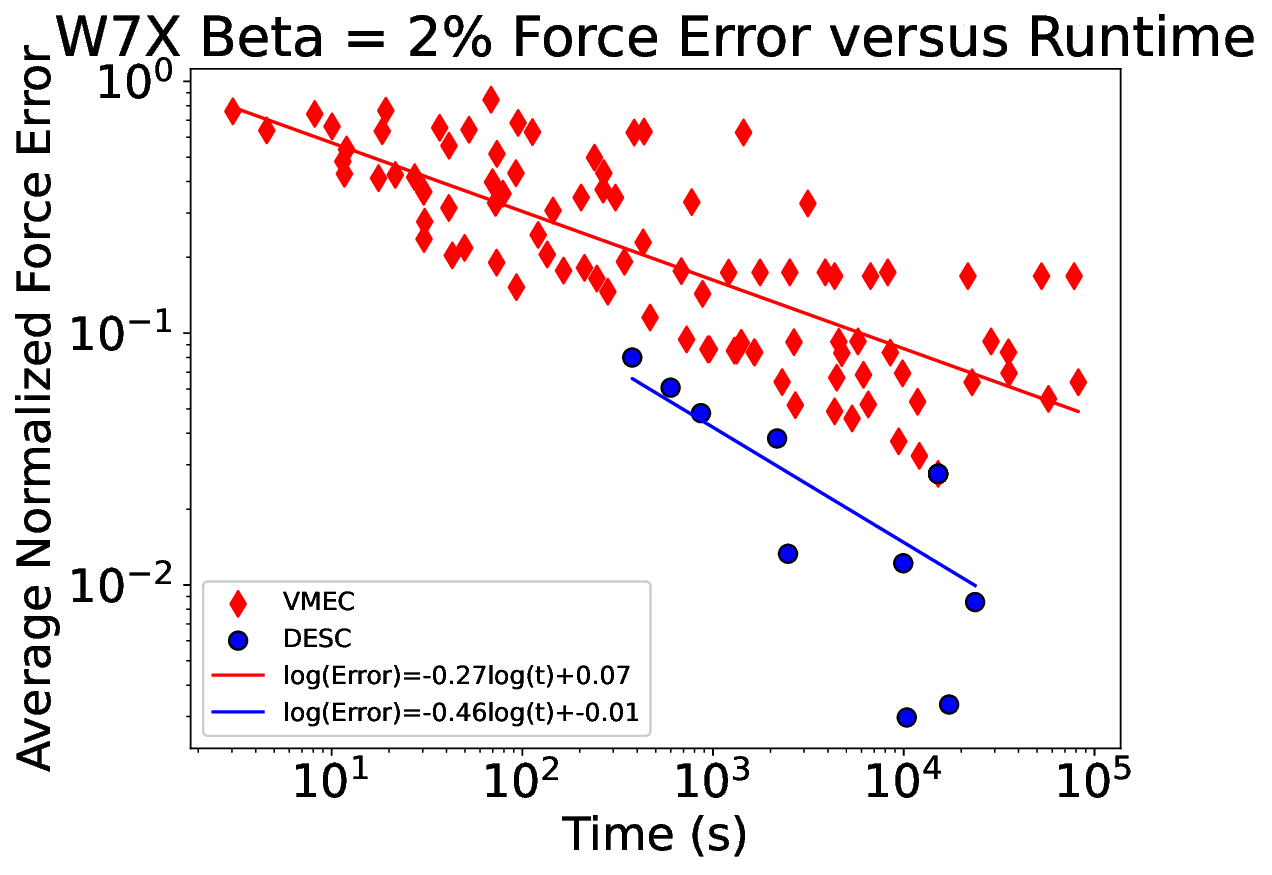}
    \caption{Scatter plot of average force error versus runtime of W7-X finite beta DESC and VMEC solutions at various resolutions, plotted along with linear fits of the results for each code. All calculations were ran on the same hardware (32GB RAM on a single AMD EPYC 7281 CPU). Note that for a given time to solution, DESC has generally an order of magnitude lower error, as seen by the best-fit lines for the results from each code.} 
    \label{fig:F_t_comp}
\end{figure}

Additionally, the average normalized force error and time-to-solution for an aggregation of a number of DESC and VMEC solutions to the W7-X-like equilibrium ran at a range of resolutions is shown in Figure \ref{fig:F_t_comp}. The resolution scan ranges are shown in Table \ref{tab:scan_parameters}. It is clear that for a given time-to-solution, DESC is able to achieve a lower average normalized force error than VMEC, indicating that DESC is able to achieve more accurate solutions than VMEC as measured by the force error metric. Often the DESC error is an order of magnitude lower than the VMEC error, as shown by the best-fit lines plotted in the figure. It should be noted that as DESC is written in Python, there is a certain amount of overhead associated with running the code versus a compiled Fortran code like VMEC. While the DESC code employs the JAX \citep{jax2018github} package, which provides just-in-time (JIT) compilation of code to improve performance, the code must first be called and compiled by JAX before the performance gains are seen. As such, the DESC solutions shown do not have runtimes lower than 2 minutes. However, pre-compilation is a planned future improvement to the JAX package, which will allow the DESC code to avoid costly JIT compilation during equilibrium solves, leading to lower initialization times. 

\begin{table}
    \centering
    \begin{tabular}{cc|c}
     & \textbf{VMEC} & \textbf{DESC} \\ \hline
      \textbf{Angular M=N}   & $[8,10,12,14,16,18,20]$ & $[8,10,12,14,16,18]$ \\ \hline
      \textbf{Radial} & $NS_{max}=[256,512,1024]$ & $L=M=N$\\\hline
      \textbf{Other} & $FTOL=[10^{-4},10^{-8},10^{-12}]$ & Index = [ANSI,Fringe]\\ \hline
    \end{tabular}
    \caption{Solution parameters scanned over in obtaining the results shown in Figure \ref{fig:F_t_comp}. Index refers to the spectral indexing scheme of the Zernike polynomials, which affects the radial resolution for a given $L$ and $M$\citep{Genberg2002,Loomis1978}}
    \label{tab:scan_parameters}
\end{table}

\subsection{Mercier Stability Comparison}\label{mercier_section}

High accuracy solutions to the MHD equilibrium equations are crucial for further analyses such as stability. This is well-known in the tokamak community, where when solving the Grad-Shafranov (GS) equation for an equilibrium reconstruction, EFIT \citep{lao_reconstruction_1985} Picard iteration errors below $10^{-4}$ \citep{xing_cake_2021,jiang_kinetic_2021} have been conventionally accepted as thresholds for reliable MHD stability analyses \citep{glasser_direct_2016,glasser_direct_2020}. These iteration error levels correspond to a similar magnitude of normalized (by the source terms) GS force error. Note that the previously shown D-shaped tokamak equilibrium solved with DESC passes the required error threshold (in terms of normalized force error) of $10^{-4}$ as shown in Figure \ref{fig:DESC_DSHAPE_Radial_conv}, while the VMEC solution fails to meet this threshold even at high radial resolution, as shown in Figure \ref{fig:VMEC_DSHAPE_Radial_conv}.

While no such rule of thumb exists for 3D equilibria in the stellarator community, VMEC's inaccuracy near the magnetic axis has been found to cause issues when conducting ideal MHD stability analyses near-axis \citep{glasser_private}. 

\begin{figure}
    \centering
    \includegraphics[keepaspectratio, width=3.2in]{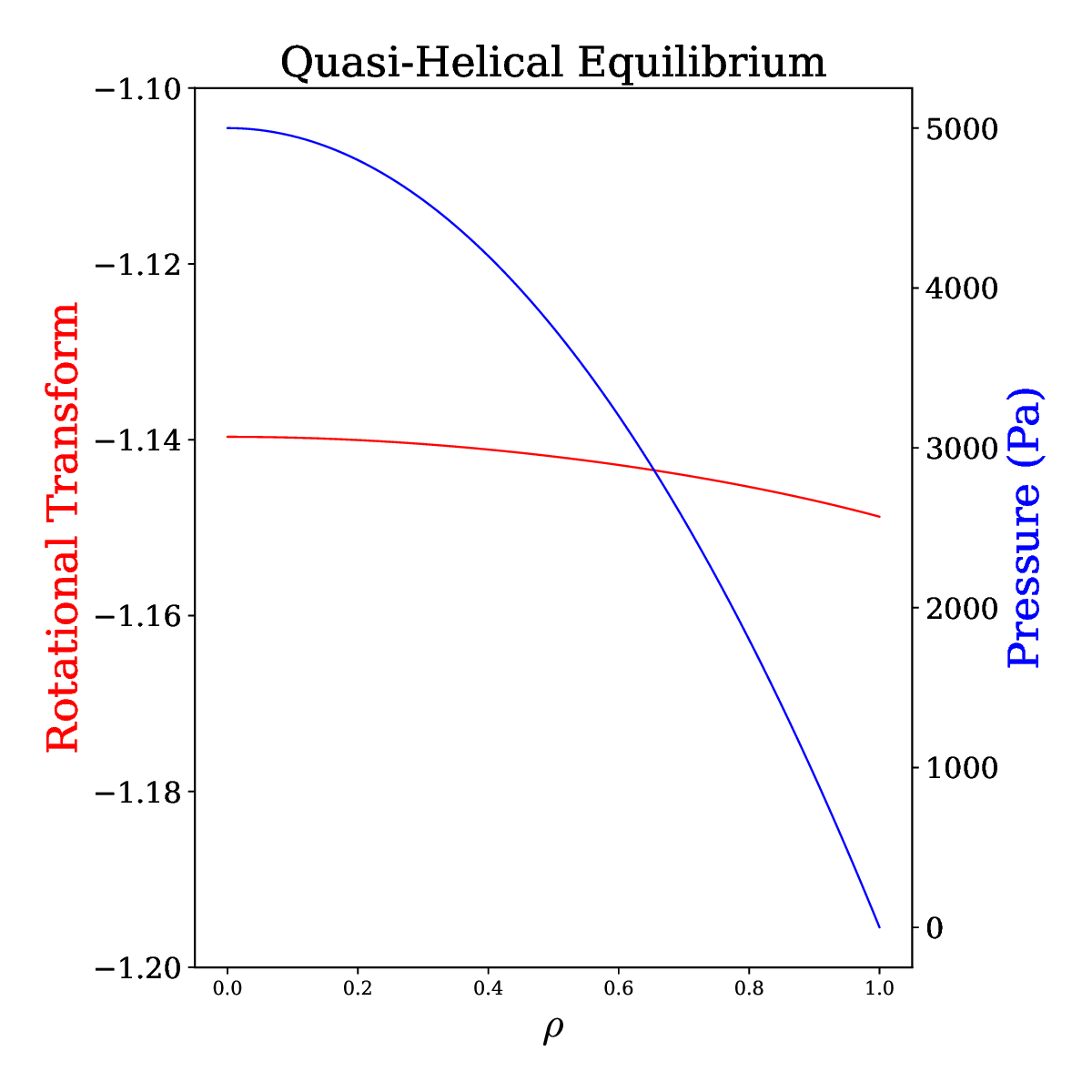}
    \caption{Pressure and rotational transform profiles of the quasi-helical equilibrium DESC solution (L=M=12, N=12).}
    \label{fig:QH_profiles}
\end{figure}

\begin{figure}
    \centering
    \includegraphics[keepaspectratio, width=1\columnwidth
    ]{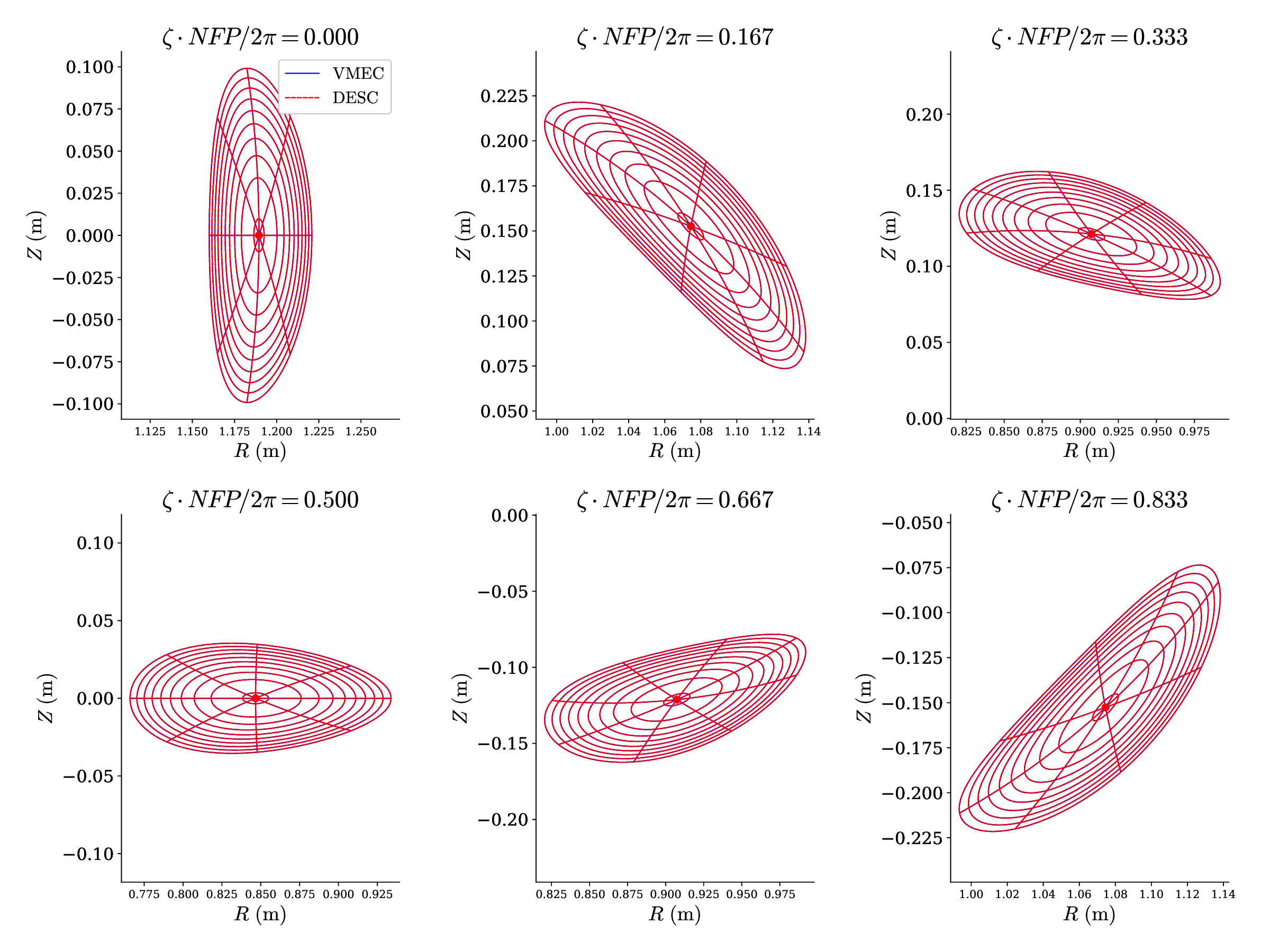}
    \caption{Flux surfaces for the VMEC (ns=801 M=N=10) and a DESC (L=M=12, N=10) quasi-helical equilibrium solution.}
    \label{fig:QH_surfaces}
\end{figure}

As an example of higher accuracy equilibria translating to more accurate stability calculations, the Mercier stability of a configuration described near eq. 4.25 of work by \citet{landreman_magnetic_2020} is calculated, with the same equilibrium solved using the DESC and VMEC codes, and compared to asymptotic formulas from near-axis expansion theory from the same work. The internal routines from DESC are used to calculate its stability metrics, while for VMEC the $D_{Merc}$ quantity that is calculated from its own internal routines and stored in its output file is used. The equilibrium is a finite beta quasi-helically symmetric configuration with an aspect ratio of $\sim20$, and was solved in VMEC at two radial resolutions $ns=[65,801]$. The same equilibrium was also solved using DESC with radial resolution $L=12$ and ANSI spectral indexing. Both codes used a poloidal resolution of $M=12$ and a toroidal resolution of $N=10$.  Instead of a fixed rotational transform profile, the equilibria were solved with a zero net toroidal current constraint in each code. The quasi-helical equilibrium used had the following pressure profile (given as a power series in $\rho = \sqrt{s} = \sqrt{\frac{\psi}{\psi_a}}$):
\begin{equation}
    p(\rho) = 5000 - 5000 \rho^2 ~ [Pa]
\end{equation}
 The pressure profile and the final rotational transform profile from the DESC equilibrium are plotted in Figure \ref{fig:QH_profiles}, and the flux surfaces are shown in Figure \ref{fig:QH_surfaces}, which shows qualitative agreement in the equilibrium solution of the two codes. However, this qualitative agreement does not necessarily lead to quantitative agreement of stability metrics, as shown by the comparison of Mercier stability in this section.
Mercier stability \citep{mercier_equilibrium_1964,mercier_MHD_book_1974} is a measure of an ideal MHD equilibrium's stability to localized interchange perturbations about a rational surface, and the criterion for stability against such perturbations can be defined as \citep{bauer_magnetohydrodynamic_1984}:
\begin{equation}
    D_{\text {Merc }}=D_{\text {Shear }}+D_{\text {Curr }}+D_{\text {Well }}+D_{\text {Geod }}>0
\end{equation}
where:
\begin{subequations}
\label{eq:mercier_full}
\begin{align}
    D_{\text {Shear }}=\frac{1}{16 \pi^2}\left(\frac{\mathrm{d} \iota}{\mathrm{d} \psi}\right)^2&\label{Dshear}\\
    D_{\text {Curr }}=-\frac{s_G}{(2 \pi)^4} \frac{\mathrm{d} \iota}{\mathrm{d} \psi}& \int \mathrm{d} S \frac{\Xi \cdot B}{|\nabla \psi|^3}\label{Dcurr}\\
    D_{\text {Well }}=\Bigg[\frac{\mu_0}{(2 \pi)^6} \frac{\mathrm{d} p}{\mathrm{~d} \psi}\Bigg(s_\psi \frac{\mathrm{d}^2 V}{\mathrm{~d} \psi^2}&\label{Dwell}\\
    -\mu_0 \frac{\mathrm{d} p}{\mathrm{~d} \psi} \int \frac{\mathrm{d} S}{B^2|\nabla \psi|} \Bigg) & \int \mathrm{d} S \frac{B^2}{|\nabla \psi|^3}\Bigg]\nonumber\\
    D_{\mathrm{Geod}}=\Bigg[\frac{1}{(2 \pi)^6}\left(\int \mathrm{d} S \frac{\mu_0 \boldsymbol{J} \cdot \boldsymbol{B}}{|\boldsymbol{\nabla} \psi|^3}\right)^2&\label{Dgeod}\\
    -\frac{1}{(2 \pi)^6}\left(\int \mathrm{d} S \frac{B^2}{|\nabla \psi|^3}\right)& \int \mathrm{d} S \frac{\left(\mu_0 \boldsymbol{J} \cdot \boldsymbol{B}\right)^2}{B^2|\boldsymbol{\nabla} \psi|^3}\Bigg]\nonumber
\end{align}
\end{subequations}


Here, $I(\psi)$ and $G(\psi)$ are the Boozer coordinate profile functions \citep{boozer_plasma_1981}, and $s_g=\pm 1$ and $s_{\psi} = \pm 1$ are the signs of $G(\psi)$ and of the normalized toroidal flux $\psi$. Furthermore, $\Xi = \mu_0 \boldsymbol{J} - I'(\psi) \boldsymbol{B}$, $V(\psi)$ is the volume enclosed by the flux surface labelled by $\psi$, and the surface integrals are taken over the given flux surface so that $dS = |\nabla \psi||\sqrt{g}|d{\vartheta}{d\varphi}$. The above equations were used to calculate Mercier stability of the equilibrium solutions from DESC and VMEC.

As a verification exercise, near-axis theory provides asymptotic expressions for the different components of Mercier stability that should match the full expressions above when $\epsilon = r/{R_{scale}}$ is small, where here $R_{scale} \sim \frac{1}{\kappa}$ is the scale length of the magnetic axis, $\kappa$ being the axis curvature, $r$ is the effective minor radius, $2\pi\psi = \pi r^2 \bar{B}$, and $\bar{B}$ is a constant reference magnetic field strength. Regardless of aspect ratio or scale length, the expansion is accurate close to the magnetic axis since $r \rightarrow 0$, and with higher aspect ratio the expansion becomes valid over a larger portion of the plasma volume.  In \citet{landreman_magnetic_2020} the asymptotic expressions for $D_{\mathrm{Geod}}$ and $D_{\text {Well}}$, the leading order (in $\epsilon$) terms of $D_{Merc}$, have been derived for quasi-symmetric equilibria:

\begin{align}
    D_{\text {Well }}&=\frac{\mu_0 p_2\left|G_0\right|}{8 \pi^4 r^2 B_0^3}\left[\frac{\mathrm{d}^2 V}{\mathrm{~d} \psi^2}-\frac{8 \pi^2 \mu_0 p_2\left|G_0\right|}{B_0^5}\right]\label{DwellQS}\\
    D_{\mathrm{Geod}}&=-\frac{2 \mu_0^2 p_2^2 G_0^4 \bar{\eta}^2}{\pi^3 r^2 B_0^{10} \iota_{N 0}^2} \int_0^{2 \pi} \mathrm{d} \varphi \frac{\bar{\eta}^4+\kappa^4 \sigma^2+\bar{\eta}^2 \kappa^2}{\bar{\eta}^4+\kappa^4\left(1+\sigma^2\right)+2 \bar{\eta}^2 \kappa^2}\label{DgeodQS}
\end{align}

where \citep{landreman_constructing_2019} $G_0$ is the zeroth order term of the Boozer profile function $G$'s expansion in a power series in effective minor radius $r$, $B_0$ is the zeroth order term in the power series expansion of the magnetic field strength $B$ in $r$,  $p_2$ is the second order term in the power series expansion of $p$ in $r$, $\bar{\eta}$ is a measure of the magnetic field strength variation, $\iota_{N0} = \iota_0 - N$ where $\iota_0$ is the zeroth order term of the expansion of $\iota$ in $r$ and $N$ is a constant integer, and $\sigma$ is a solution to the ordinary differential equation defined in Eq. 2.14 of \citet{landreman_constructing_2019}.





Thus to leading order, the Mercier stability of a quasi-symmetric stellarator calculated by the full expressions in Eqs. \eqref{eq:mercier_full} should agree with the sum of Eqs. \eqref{DwellQS} and \eqref{DgeodQS} at small $\epsilon$. Any deviation from agreement in a high-aspect-ratio equilibrium, especially near the magnetic axis, would then indicate an inaccuracy in the underlying equilibrium, as the calculated stability fails to agree with the asymptotic expression in the region where it is most valid.

\begin{figure}
    \centering
    \includegraphics[keepaspectratio, width=3.25in]{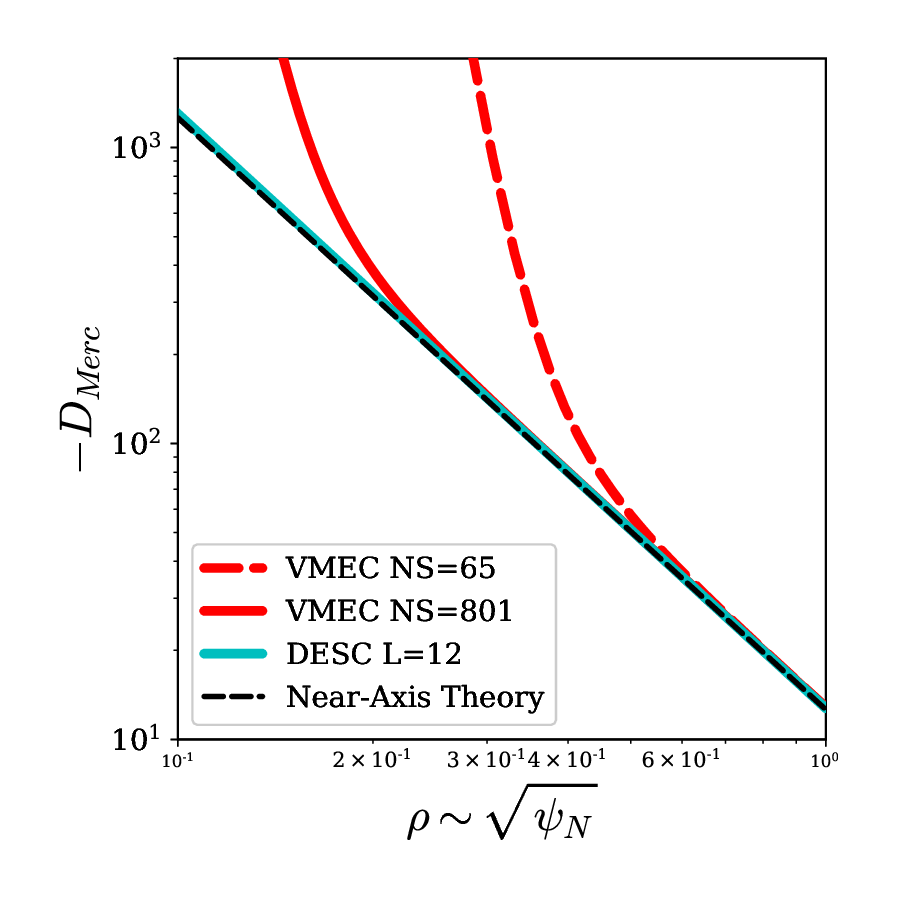}
    \caption{Mercier stability calculated from VMEC equilibria of increasing radial resolution, as compared to a DESC equilibrium of L=12. Both codes were ran with toroidal resolution of N=10 and poloidal resolution of M=12. The DESC solution compares much better with the asymptotic value of $D_{Merc}$ near-axis, while the VMEC solution even with high resolution fails to resolve the stability near-axis.} 
    \label{fig:D_merc_comp}
\end{figure}

\begin{figure}
    \centering
    \includegraphics[keepaspectratio, width=3.25in]{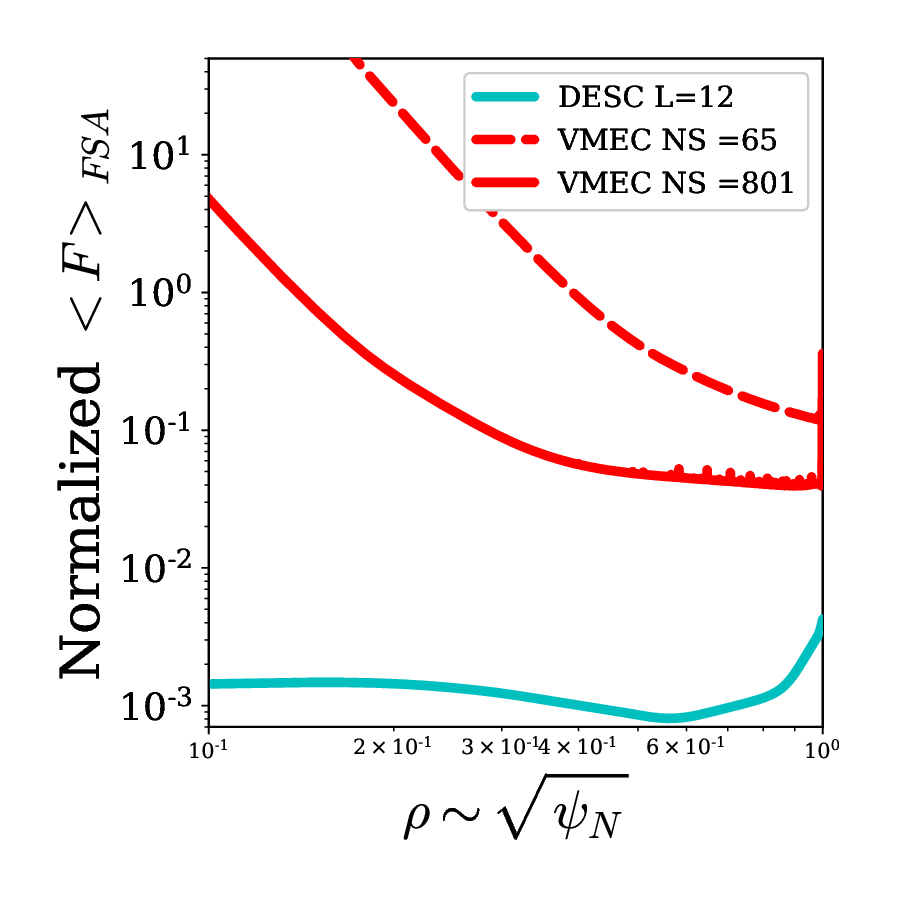}
    \caption{Normalized force error flux surface average of the VMEC and DESC equilibria corresponding to the calculations in Figure \ref{fig:D_merc_comp}.} 
    \label{fig:F_fsa_dmerc_comp}
\end{figure}

Figure \ref{fig:D_merc_comp} shows $D_{Merc}$ (negated for the sake of plotting on a log scale, as the equilibrium is Mercier-unstable) of the two equilibria computed with VMEC in red, while that calculated from the DESC equilibrium is in cyan. Plotted as well is the value of $D_{Merc}$ given by the asymptotic expressions. It can be seen that the stability computed from the DESC equilibrium agrees well with the asymptotic expression across the entirety of the plasma volume, but most importantly near the magnetic axis, indicating the equilibrium is sufficiently well-resolved for accurate Mercier stability calculation. The VMEC equilibrium requires high radial resolution to agree with the asymptotic expression inside of $\rho < 0.5$, and even then differs from the asymptotic value nearer to the magnetic axis, indicating a failure to accurately satisfy the ideal MHD equilibrium equations there. 
Upon inspection of the individual terms in the $D_{Merc}$ sum, the $D_{Geod}$ term is the most significant part of $D_{Merc}$ in the core of the equilibrium, and it is this dominant term which is responsible for the large deviation in $D_{Merc}$ from the asymptotically-derived values near the axis. The lack of convergence of $D_{Geod}$ in VMEC with increasing resolution has been seen by previous work \citep{landreman_magnetic_2020}. This deviation is roughly correlated with the radial position where the force error in each VMEC solution begins to increase sharply nearing the axis, shown in Figure \ref{fig:F_fsa_dmerc_comp}.




One possible explanation for such a difference in accuracy between the DESC and VMEC solutions is that DESC's 3D spectral basis results in more accurate radial derivatives as compared to VMEC's finite differencing. Of the two leading-order terms in $D_{Merc}$ near the axis, the $D_{Geod}$ term requires higher order radial derivatives of the coordinates $R,Z$, due to the presence of the current density $\mathbf{J}$ in the expression. So, an inaccuracy in the radial derivatives would most strongly affect this term, leading to the lack of agreement with the asymptotic result at the axis. This inaccuracy in radial derivatives also manifests itself in worse force error.
This shows the importance of accurate equilibria, especially near the axis, as VMEC's lack of accuracy leads to disagreement with asymptotic theory in the region where the theory is most valid, while DESC's accurate treatment of the axis results in accurate Mercier stability calculations as well.

\section{Conclusions}
In conclusion, a comparison of the VMEC and DESC codes was carried out. DESC was shown to have more accurate equilibrium solutions than VMEC as measured by force balance error, and to have faster runtimes for a given solution accuracy. DESC was also shown to have improved radial convergence as compared to VMEC, owing to its spectral basis in all three coordinates. Further, inaccuracies of VMEC solutions near the axis were seen, which could be tied to the unphysical modes in the VMEC Fourier spectrum that do not scale correctly with radius near the axis. DESC solutions, on the other hand, do not have this problem, and were shown to be accurate near the axis. Solution accuracy is necessary in order to accurately calculate stability metrics, and this is explicitly shown for a Mercier stability calculation, where the DESC solution is found to better agree with the asymptotic expansion near-axis than the VMEC solution. Future plans for development of the DESC code with regards to computation speed include implementing MPI parallelization to better take advantage of CPUs and benchmarking the code's GPU capabilities. Pre-compilation of functions with JAX is also foreseen, which should aid in reducing initialization times. Additionally, further DESC verification can include comparing results to other equilibrium codes such as SPEC, and the effect of solution accuracy can be investigated in other metrics such as fast particle confinement.\\
\textbf{Acknowledgements}
The authors gratefully acknowledge helpful discussions with Matt Landreman concerning near-axis expansion theory and Mercier stability.\\
\textbf{Funding}
This work was supported by the U.S. Department of Energy under contract numbers DE-AC02-09CH11466, DE-SC0022005 and Field Work Proposal No. 1019. The United States Government retains a non-exclusive, paid-up, irrevocable, world-wide license to publish or reproduce the published form of this manuscript, or allow others to do so, for United States Government purposes.\\
\textbf{Data Availability Statement}
The data and scripts used in this manuscript are freely available on Zenodo \citep{panici_dario_2022_6539680} at \url{https://doi.org/10.5281/zenodo.6539680}.\\
\textbf{Declaration of Interests}
The authors report no conflict of interest.
\bibliographystyle{jpp}
\bibliography{bib.bib}

\appendix

\section{Force Balance Error from VMEC \texorpdfstring{$R,Z,\lambda$}{VMEC outputs} Fourier Coefficients} \label{appendix:F_bal_deriv_VMEC}

With the cylindrical coordinate system $\mathbf{x} = (R,\phi,Z)$.
VMEC uses as its computational coordinates $\mathbf{\alpha} = (s,u,v)$, with $s$ being a radial coordinate proportional to the normalized toroidal flux, $u$ a poloidal-like angle, and $v$ is the geometric toroidal angle for one field period:

\begin{subequations}
\begin{equation}
    s = \frac{\psi}{\psi_a} \hspace{2mm}, \hspace{2mm} 0\leq s \leq 1
\end{equation}
\begin{equation}
    u = \theta^* - \lambda(s,u,v)\hspace{2mm}, \hspace{2mm} 0\leq u \leq 2\pi
\end{equation}
\begin{equation}
    v = \phi  \hspace{2mm}, \hspace{2mm} 0\leq v \leq \frac{2\pi}{N_{FP}}
\end{equation}
\end{subequations}

where $\psi_a$ is the toroidal flux enclosed by the plasma boundary (at s=1) that is normalized by $2\pi$, $\lambda(s,u,v)$ is a function periodic in $(u,v)$ that converts $u$ to a straight-field-line poloidal angle $\theta^*$, $N_{FP}$ is the number of field periods in the device, $\phi$ is the usual geometric toroidal angle coordinate. 
\\

The covariant basis vectors $\mathbf{e}_{i} = \frac{\partial \mathbf{x}}{\partial \alpha_i}$ for the $\alpha=(s,u,v)$ coordinate system are:

\begin{subequations}
\begin{equation}
    \es = \begin{bmatrix} \Rs \\ 0 \\ \Zs \end{bmatrix}
\end{equation}
\begin{equation}
    \eu = \begin{bmatrix} \Ru \\ 0 \\ \Zu \end{bmatrix}
\end{equation}
\begin{equation}
    \ev = \begin{bmatrix} \Rv \\ R \\ \Zv \end{bmatrix}
\end{equation}
\end{subequations}
and the notation ${\mathbf e}_{\alpha\gamma}$ is used as a shorthand for $\partial_\gamma \left( {\mathbf e}_{\alpha} \right)$.
The Jacobian and its partial derivatives are calculated from the basis vectors as  
\begin{subequations}
\begin{align}
\sqrt{g} &= \es\cdot\eu\times\ev \\
\partial_s \left(\sqrt{g}\right) &= \ess\cdot\eu\times\ev + \es\cdot\eus\times\ev + \es\cdot\eu\times\evs \\
\partial_u \left(\sqrt{g}\right) &= \esu\cdot\eu\times\ev + \es\cdot\euu\times\ev + \es\cdot\eu\times\evu \\
\partial_v \left(\sqrt{g}\right) &= \esv\cdot\eu\times\ev + \es\cdot\euv\times\ev + \es\cdot\eu\times\evv
\end{align}
\end{subequations}

Contravariant basis vectors  $\mathbf{e}^{i} = \nabla\alpha_i$ are given by:

\begin{subequations}
\begin{align}
    \eS &= \frac{\eu \times \ev}{\sqrt{g}}\\
    \eU &= \frac{\ev \times \es}{\sqrt{g}}\\
    \eV &= \frac{\es \times \eu}{\sqrt{g}}\\
\end{align}
\end{subequations}

The metric tensor components are given by:

\begin{subequations}
\begin{align}
g^{ss} &= \eS\cdot\eS \\
g^{vv} &= \eV\cdot\eV \\
g^{uu} &= \eU\cdot\eU \\
g^{uv} &= \eU\cdot\eV.
\end{align}
\end{subequations}

Recall that the magnetic field can be written in the form 
\begin{subequations}
\begin{align}
\mathbf{B} &= B_s \eS + B_u \eU + B_v \eV \\
           &= B^u \eu + B^v \ev
\end{align}
\end{subequations}
and that in the VMEC coordinate system the contravariant components of the field are given by \citep[p.~3]{hirshman_steepestdescent_1983}:
\begin{subequations}
\begin{align}
B^u &= \frac{1}{\sqrt{g}} \Big(\chi' - \psi' \lv \Big)\\
B^v &= \frac{1}{\sqrt{g}} \psi'\Big(1 + \lu \Big)
\end{align}
\end{subequations}

where $2\pi \chi(s)$ and $2\pi \psi(s)$ are the poloidal and toroidal magnetic fluxes, respectively, the prime denotes a radial derivative $\partial / \partial s$, and $\lambda$ is a function periodic in $u,v$ with zero average over a magnetic surface, $\iint dudv\lambda = 0$. 
\\
\\
MHD force balance equilibrium is given by:
\begin{subequations}

\begin{align}
    \mathbf{F} = -\mathbf{J} \times \mathbf{B} + \nabla p &= 0 \label{F_bal}\\
    \nabla \times \mathbf{B}& = \mu_0 \mathbf{J}\\
    \nabla \cdot \mathbf{B} &= 0
\end{align}
\end{subequations}

The magnetic field can be written as:

\begin{subequations}
\begin{align} \label{Bfield}
   \mathbf{B} &= \nabla v \times \nabla \chi + \nabla \psi \times \nabla \theta^*\\
   &= B^u \eu + B^v \ev
\end{align}
\end{subequations}

where $\theta^* = u + \lambda(s,u,v)$ is a straight field line poloidal angle. Inserting Eq.\eqref{Bfield} into Eq.\eqref{F_bal} yields:

\begin{equation}
    \mathbf{F} = F_s \nabla s + F_{\beta} \mathbf{\beta}
\end{equation}
where
\begin{subequations}
\begin{align}
    F_s &= \sqrt{g}(J^vB^u - J^uB^v) + p'\\
    F_{\beta} &= J^s
\end{align}
\end{subequations}
where $\beta = \sqrt{g}(B^v \nabla u - B^u \nabla v)$ and $J^i = \mathbf{J} \cdot \nabla \alpha_i = \mu_0^{-1} \nabla \cdot (\mathbf{B} \times \nabla \alpha_i)$

Contravariant components of $\mathbf{J}$ can be written with the derivatives of covariant B components $\mathbf{B} = B_i e^i$:
\begin{subequations}
\begin{align}
    J^s &= \mu_0^{-1} \nabla \cdot (\mathbf{B} \times \nabla s) = \frac{1}{\mu_0 \sqrt{g}} \nabla \cdot (B_v \nabla u - B_u \nabla v)\\
    J^s&= \frac{1}{\mu_0 \sqrt{g}}\Big(\frac{\partial B_v}{\partial u} - \frac{\partial B_u}{\partial v}\Big) = F_{\beta}
\end{align}
\begin{align}
    J^u &= \mu_0^{-1} \nabla \cdot (\mathbf{B} \times \nabla u) = \frac{1}{\mu_0 \sqrt{g}} \nabla \cdot (-B_v \nabla s + B_s \nabla v)\\
    J^u&= \frac{1}{\mu_0 \sqrt{g}}\Big(\frac{\partial B_s}{\partial v} - \frac{\partial B_v}{\partial s}\Big) 
\end{align}
\begin{align}
J^v &= \mu_0^{-1} \nabla \cdot (\mathbf{B} \times \nabla v) = \frac{1}{\mu_0 \sqrt{g}} \nabla \cdot (B_u \nabla s - B_s \nabla u)\\
   J^v &= \frac{1}{\mu_0 \sqrt{g}}\Big(\frac{\partial B_u}{\partial s} - \frac{\partial B_s}{\partial u}\Big) 
\end{align}
\end{subequations}

The covariant components of B are:

\begin{subequations}\label{covB}
\begin{align}
    B_s &= \mathbf{B} \cdot \es = (B^u \eu + B^v \ev) \cdot \es\\
    B_u &= \mathbf{B} \cdot \eu = (B^u \eu + B^v \ev) \cdot \eu\\
    B_v &= \mathbf{B} \cdot \ev = (B^u \eu + B^v \ev) \cdot \ev
\end{align}
\end{subequations}

The partial derivatives of the contravariant components of B are for $B^u$:
\begin{subequations}
\begin{align}
\partial_s B^u &=  -\frac{\partial_s(\sqrt{g})}{g} \left(\chi'-\psi'\lv\right) + \frac{1}{\sqrt{g}} \left(\chi'' - \psi''\lv - \psi' \lsv\right)\\
\partial_u B^u &=  -\frac{\partial_u(\sqrt{g})}{g} \left(\chi'-\psi'\lv\right) + \frac{1}{\sqrt{g}} \left(- \psi' \luv\right)\\
\partial_v B^u &=  -\frac{\partial_v(\sqrt{g})}{g} \left(\chi'-\psi'\lv \right) + \frac{1}{\sqrt{g}} \left(- \psi' \lvv \right)\\
\end{align}
\end{subequations}

and for $B^v$:
\begin{subequations}
\begin{align}
\partial_s B^v &=  \left(-\frac{\partial_s(\sqrt{g})}{g} \psi' + \frac{\psi''}{\sqrt{g}}\right)\left(1+\lu\right) + \frac{\psi'}{\sqrt{g}}\left(\lsu\right)\\
\partial_u B^v &=  -\frac{\partial_u(\sqrt{g})}{g} \psi' \left(1+\lu\right) + \frac{\psi'}{\sqrt{g}}\left(\luu\right)\\
\partial_v B^v &=  -\frac{\partial_v(\sqrt{g})}{g} \psi' \left(1+\lu\right) + \frac{\psi'}{\sqrt{g}}\left(\luv\right)\\
\end{align}
\end{subequations}

With these defined, and using Eq.\eqref{covB}, the partial derivatives of the covariant components of B are then:

\begin{subequations}
\begin{align}
    \covBpartial{s}{u}
\end{align}
\begin{align}
    \covBpartial{s}{v}
\end{align}
\begin{align}
    \covBpartial{u}{s}
\end{align}
\begin{align}
    \covBpartial{u}{v}
\end{align}
\begin{align}
    \covBpartial{v}{s}
\end{align}
\begin{align}
    \covBpartial{v}{u}
\end{align}

\end{subequations}

With these, all of the required derivatives to evaluate the force components $F_s$ and $F_\beta$ are known. The magnitudes of the directions of each component are:

\begin{subequations}
\begin{align}
    ||\nabla s||_2 &= \sqrt{\eS \cdot \eS} = \sqrt{g^{ss}}\\
    ||\mathbf{\beta}||_2 &=
    ||\sqrt{g}\left(B^v\eU - B^u \eV\right)||_2\\
    &= \sqrt{g}\sqrt{(B^v)^2g^{uu} + (B^u)^2g^{vv} - 2B^vB^ug^{uv}} 
\end{align}
\end{subequations}

The magnitude of force balance error is then:

\begin{equation}
    ||\mathbf{F}||_2 = \sqrt{\left(F_s\right)^2g^{ss} + \left(F_{\beta}\right)^2 \left(||\mathbf{\beta}||_2\right)^2}
\end{equation}

$R, Z,~\text{and } \lambda$ are explicitly known analytically only in $u,v$ on discrete flux surfaces on the radial grid $s$. $\lambda$ also is calculated on a half mesh offset from the main radial grid, and must be interpolated onto the main radial grid first. So, numerical derivatives are used for all of the radial derivatives $\partial_s$ of $R, Z,~\text{and } \lambda$ necessary to calculate $\bm{F}$. The derivatives are carried out on the Fourier coefficients RMNC,ZMNS, and LMNS (to get the Fourier coefficients of the derivatives, e.g. $\frac{\partial RMNC}{\partial s}|_{s_i} = RSMNC = \frac{RMNC(s_{i+1}) -RMNC(s_{i})}{\Delta s} $).

\section{VMEC Force Error Spikes}
\label{appendix:spikes}

In the W7-X finite beta equilibria computed here, spikes are observed in the calculated force error flux surface average.

\begin{figure}
    \centering
    \includegraphics[keepaspectratio, width=3.25in]{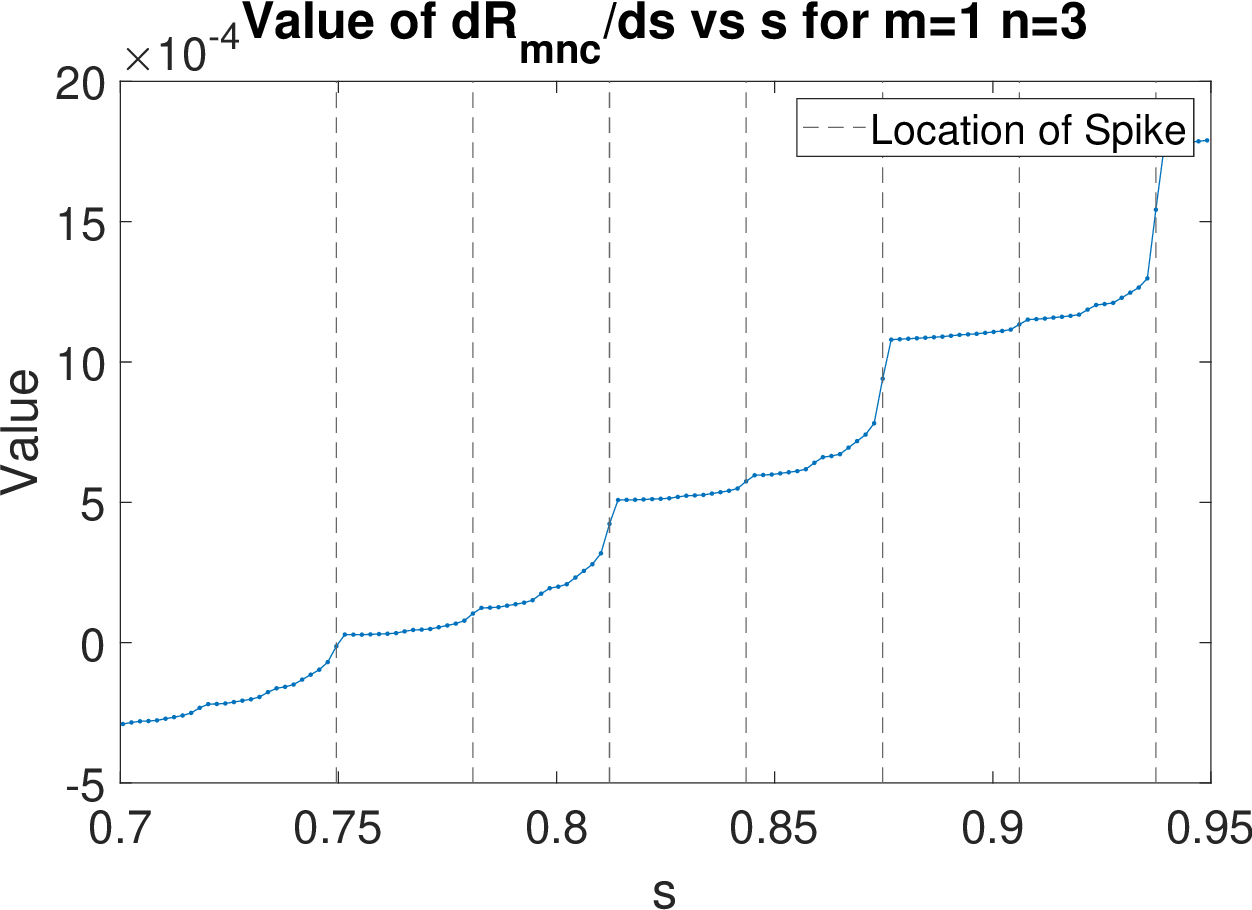}
    \caption{RMNC m=3 n=1 coefficient's first radial derivative (found with finite differences) for W7-X M=N=16 with ns=512.}
    \label{fig:RSMNC_spike}
\end{figure}

\begin{figure}
    \centering
    \includegraphics[keepaspectratio, width=3.25in]{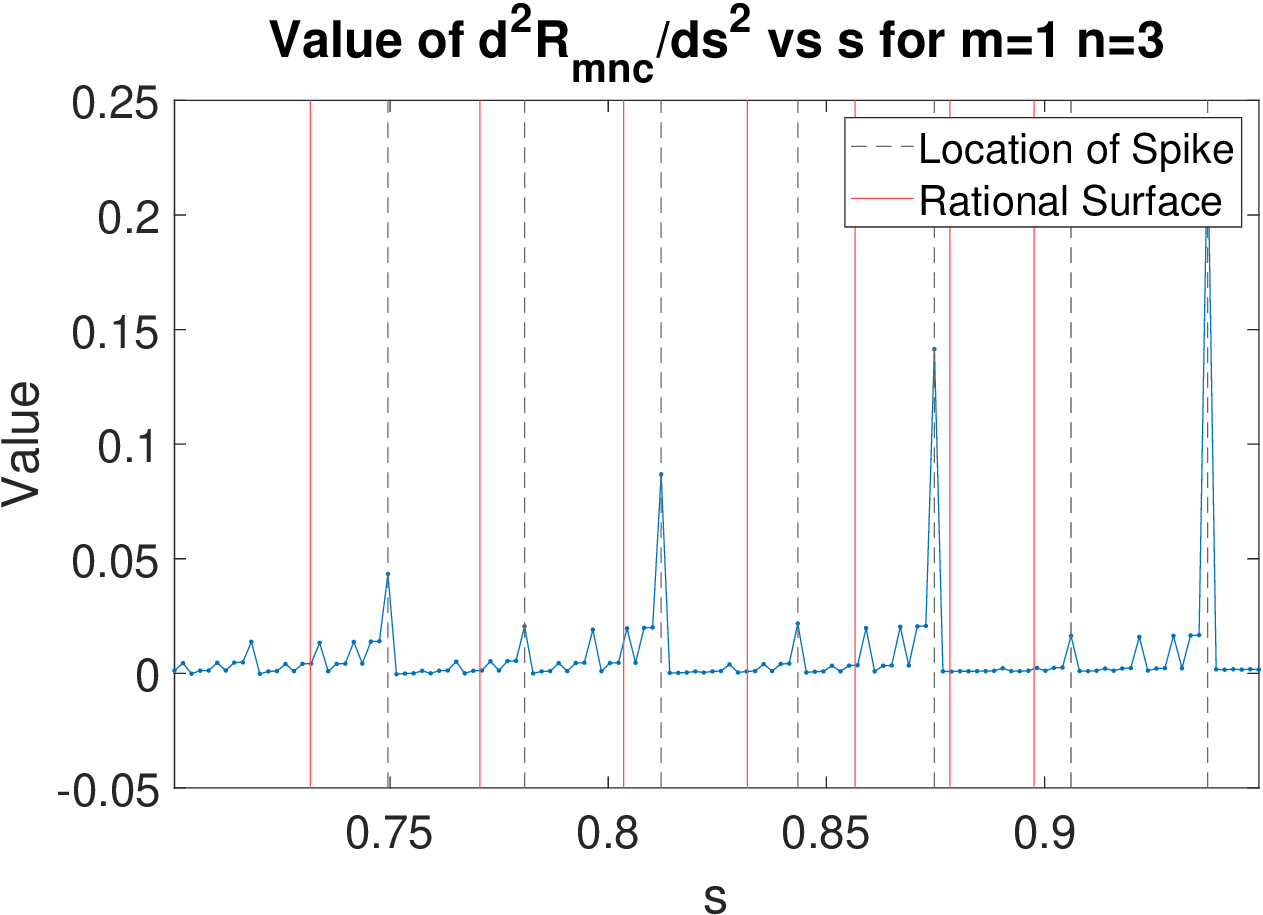}
    \caption{RMNC m=3 n=1 coefficient's second radial derivative (found with finite differences) for W7-X M=N=16 with ns=512.}
    \label{fig:RSSMNC_spike}
\end{figure}

These spikes correspond to discontinuous jumps in the radial derivatives of the fourier coefficients for $R,Z$, shown in Figures \ref{fig:RSMNC_spike} and \ref{fig:RSSMNC_spike} for the $m=3~,n=1$ mode of $R_{mnc}$ (chosen only as a representative example, this is seen in other mode numbers as well). Note that these spikes do not correspond to any low-order rationals, plotted in Figure \ref{fig:RSSMNC_spike}, so they do not stem from current singularities at rational surfaces. This is further supported by the parallel current density not exhibiting singular behavior at the rational surfaces plotted, shown in Figure \ref{fig:Jparallel_spikes}. Plotted are all rationals $N/M$ in the ranges $M=(1,19),~N=(1,18)$ that lie in the iota profile (the profile is plotted in Figure \ref{fig:W7X_profiles}).  

\begin{figure}
    \centering
    \includegraphics[keepaspectratio,width=3in]{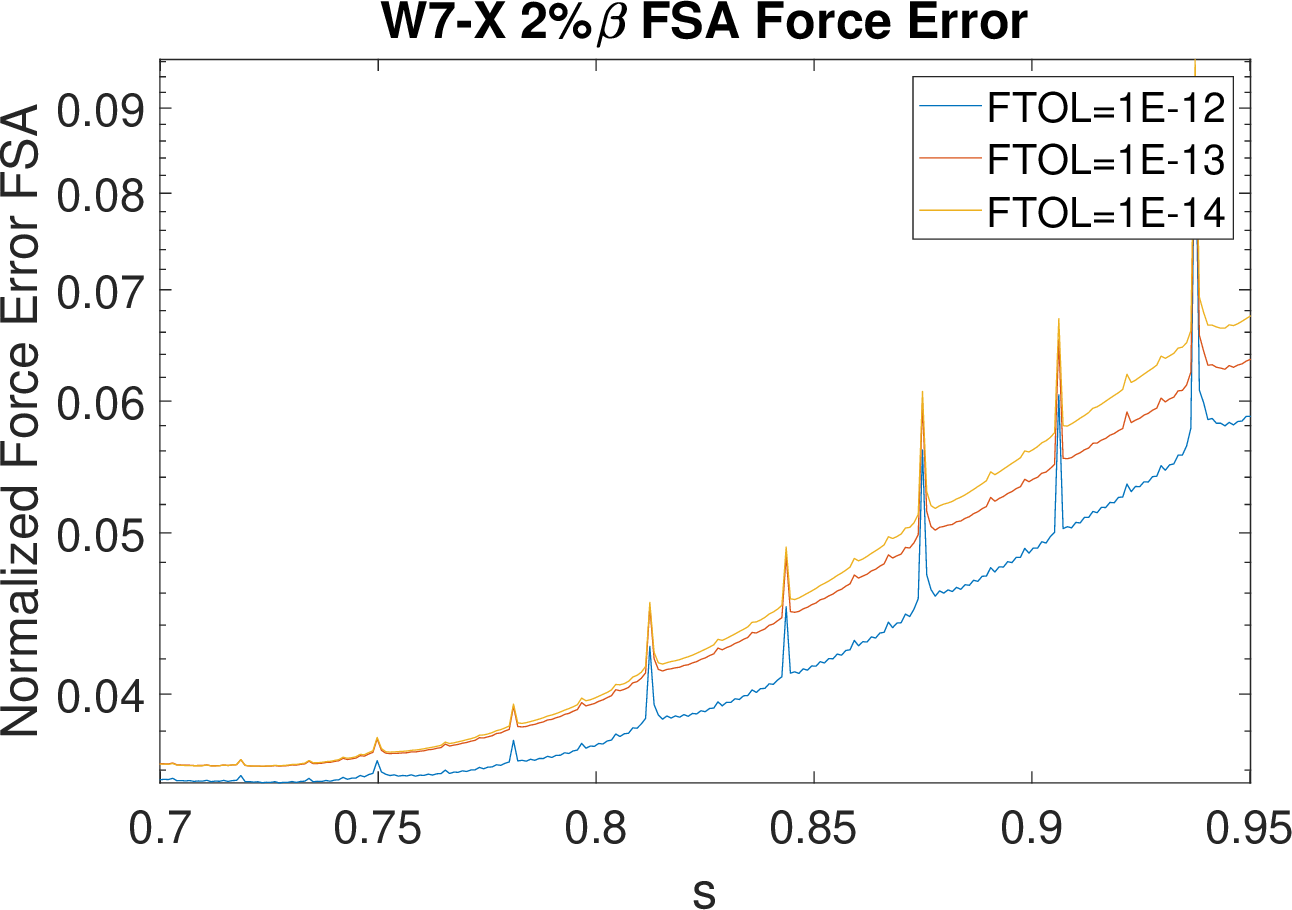}
    \caption{W7-X flux surface average of normalized force error versus $\rho$ for increasingly tighter solver tolerance (all with angular resolution of M=N=16 and NS=1024 flux surfaces). 2nd order finite differences were used as the radial derivative in calculating the force error.}
    \label{fig:ftol_scan}
\end{figure}

\begin{figure}
    \centering
    \includegraphics[keepaspectratio, width=3.25in]{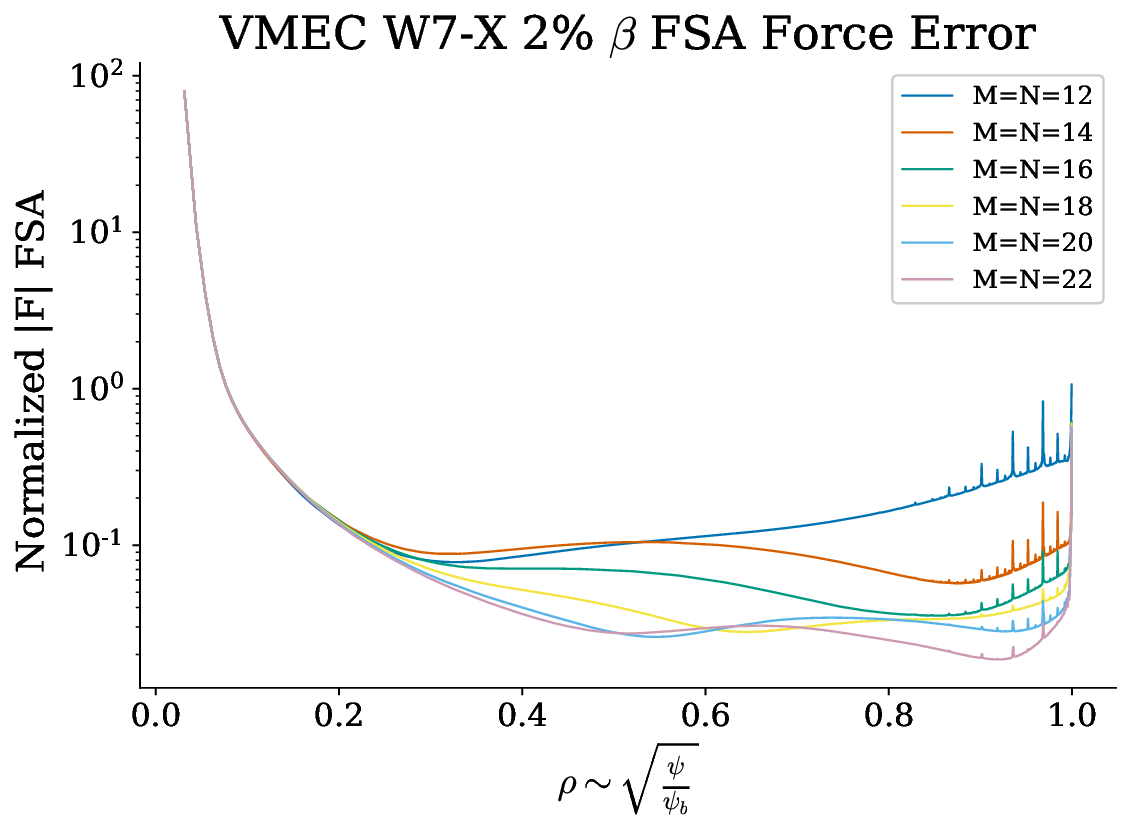}
    \caption{W7-X flux surface average of normalized force error versus $\rho$ for increasing VMEC angular resolution (all with radial resolution of $NS=1024$). 2nd order finite differences were used as the radial derivative in calculating the force error.}
    \label{fig:F_fsa_MN_conv}
\end{figure}

Running the same equilibrium with higher solver tolerances (shown in Figure \ref{fig:ftol_scan}), and with higher angular resolution, such as in Figure \ref{fig:F_fsa_MN_conv}, do not completely eliminate these spikes. Increasing the FTOL paramater past $1E-14$ resulted in the equilibrium solve taking prohibitively long (longer than 24 hours when ran with 32GB RAM on a single AMD EPYC 7281 CPU), so the tolerance scan at NS=1024, M=N=16 was not carried out past $FTOL=1E-14$.

\section{VMEC and DESC Convergence Scans}

\begin{figure}
    \centering
    \includegraphics[keepaspectratio, width=3.25in]{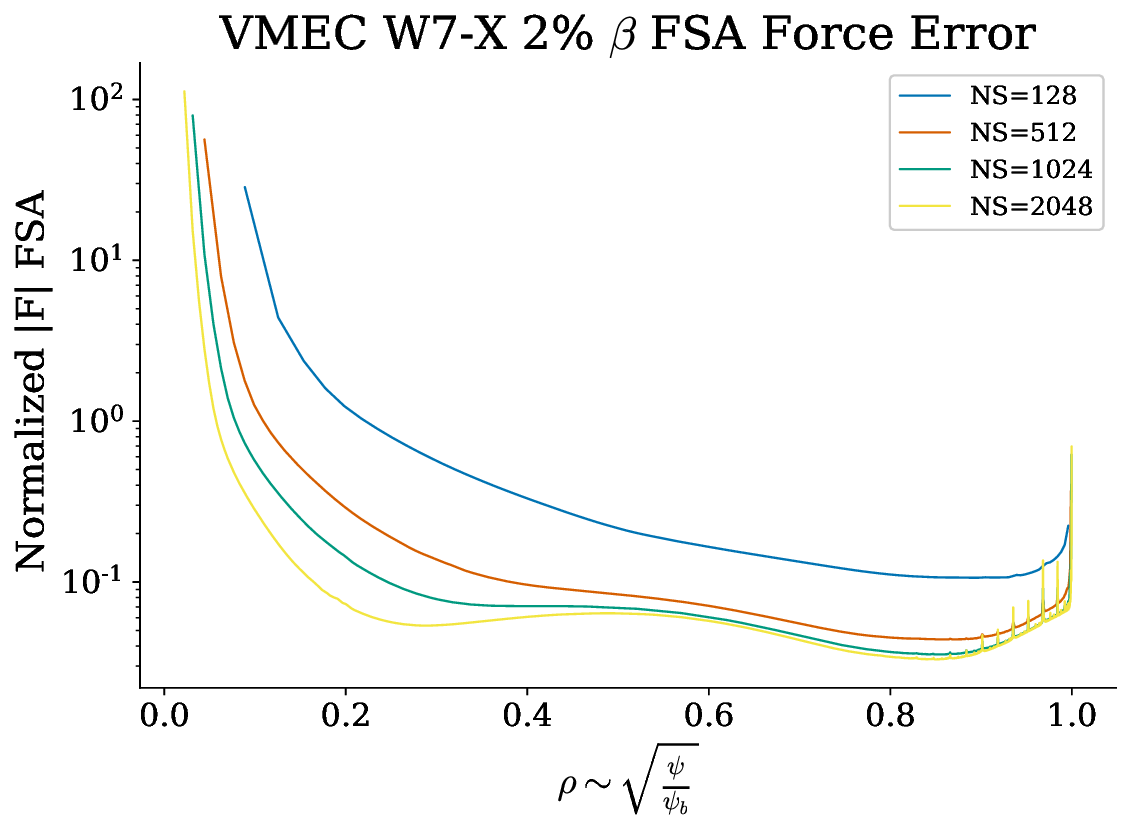}
    \caption{W7-X flux surface average of normalized force error versus $\rho$ for increasing VMEC radial resolution (all with angular resolution of M=N=16). The force error does not decrease appreciably past 1024 surfaces for most of the plasma volume, and the error spikes near the edge increase in size as NS increases. 2nd order finite differences were used as the radial derivative in calculating the force error.}
    \label{fig:F_fsa_ns_conv}
\end{figure}

\begin{figure}
    \centering
    \includegraphics[keepaspectratio, width=3.1in]{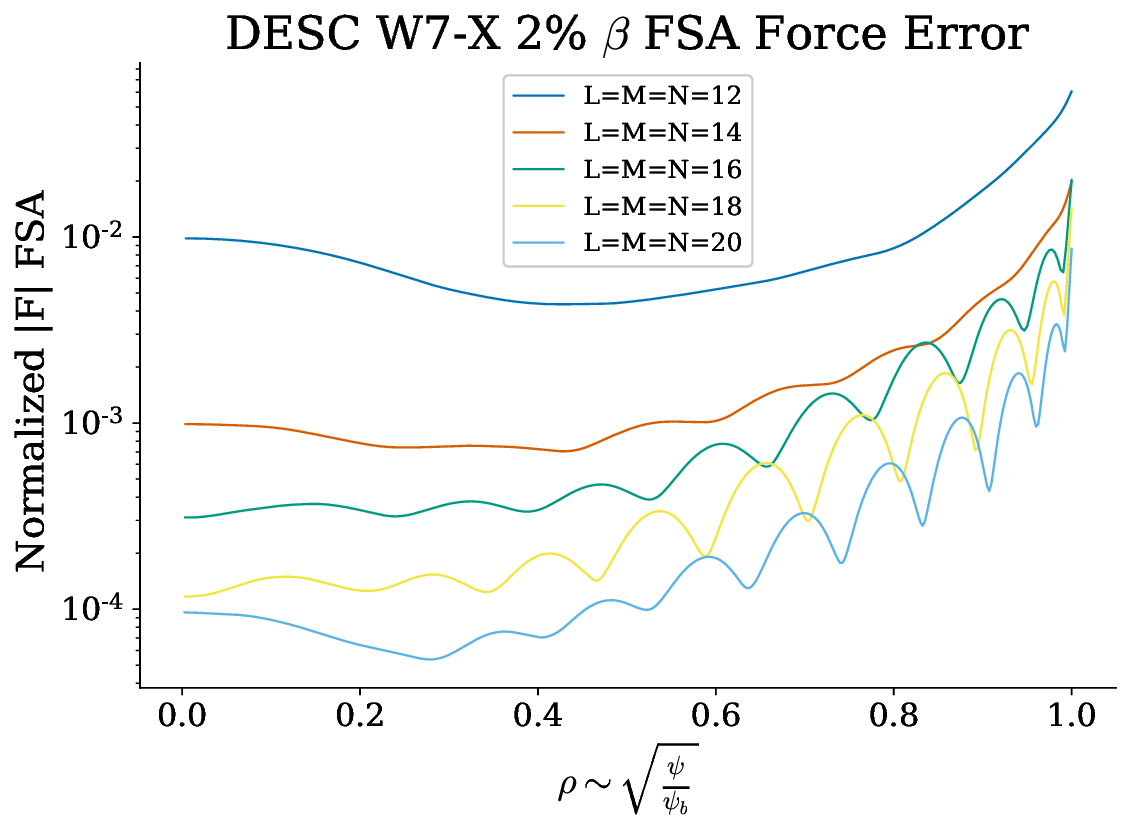}
    \caption{W7-X flux surface average of normalized force error versus $\rho$ for increasing DESC angular and radial resolution. The ANSI Zernike indexing pattern was used \citep{dudt_desc_2020}.}
    \label{fig:F_fsa_DESC_MN_conv}
\end{figure}

In computing solutions for the comparison in this paper, convergence scans were carried out with each code, the results of which are compiled here. In Figure \ref{fig:F_fsa_ns_conv}, the results for running VMEC at an angular resolution of $M=N=16$ (512 Fourier modes per surface) for increasing number of radial surfaces are shown. It can be seen that after $NS=1024$, the normalized force error does not decrease appreciably across most of the volume, and the spikes in error near the edge become much more pronounced with the $NS=2048$.\\
Next, a scan over angular resolution was carried out in VMEC, and shown in Figure \ref{fig:F_fsa_MN_conv}. The force error is seen to decrease with increasing angular resolution across the whole volume until $M=N=20$, where it begins to stagnate and not uniformly decrease. A similar scan was carried out in DESC, and shown in Figure \ref{fig:F_fsa_DESC_MN_conv}. In DESC, due to the Fourier-Zernike basis, the poloidal and radial modes are coupled, so increasing the poloidal resolution $M$ also increases the radial resolution $L$. It can be seen that around $L=M=N=16$, the normalized force error begins to not decrease uniformly across the plasma volume. The minima in the force error flux surface averages here correspond to collocation points.

\begin{figure}
    \centering
    \includegraphics[keepaspectratio, width=3.7in]{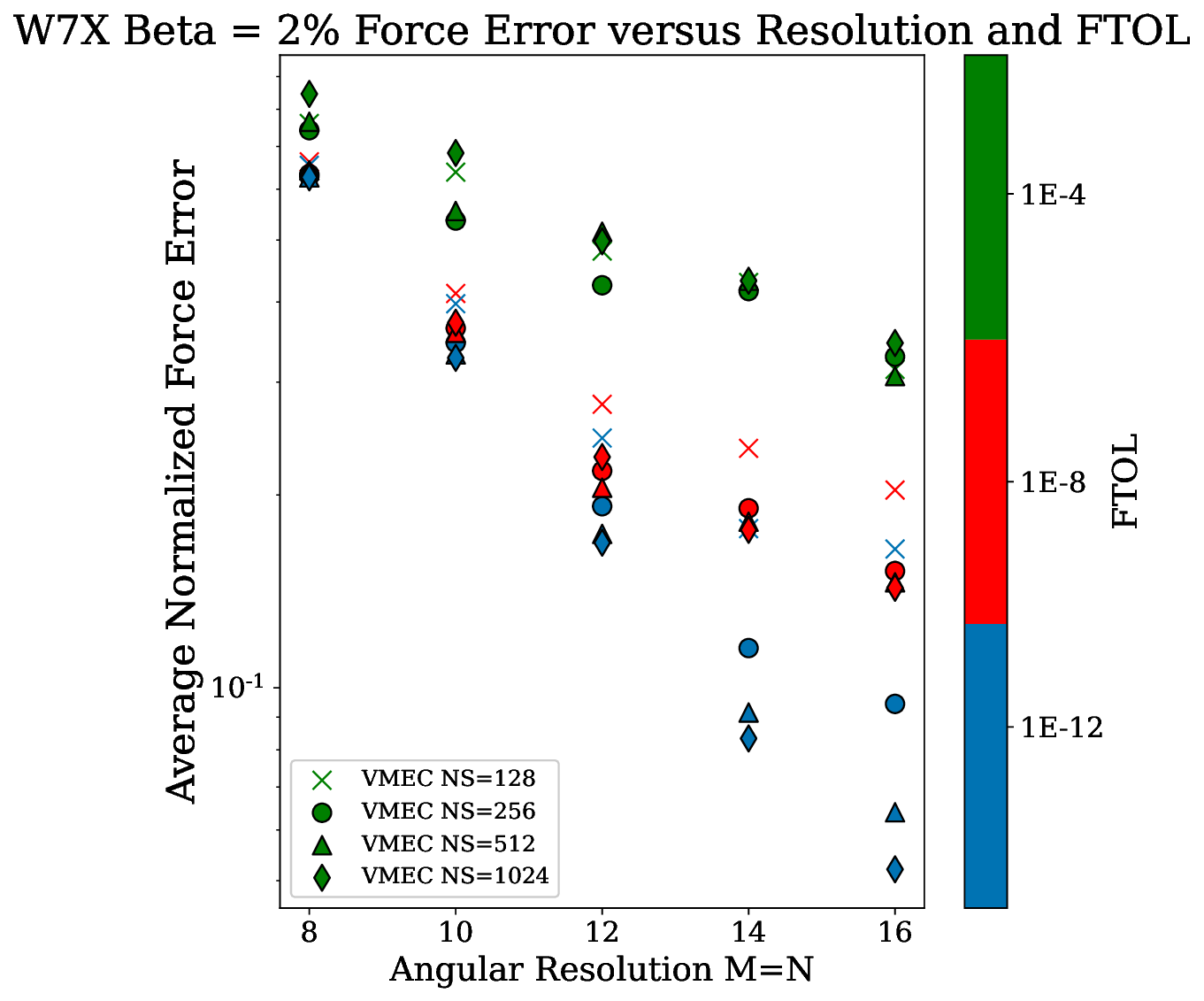}
    \caption{VMEC results labelled with ftol, showing that low ftol results in stagnation in error decrease with increasing resolution, as expected}
    \label{fig:F_res_ftol}
\end{figure}

Among the parameters scanned over for the VMEC solutions shown in this paper was the FTOL solver tolerance parameter. Shown in Figure \ref{fig:F_res_ftol} are results of running the W7-X-like equilibrium at different angular and radial resolutions, and at a range of FTOL values. It can be seen that at low angular resolutions (lower than the boundary Fourier series resolution of $M=N=12$), the FTOL parameter does not affect the solution accuracy much. This is likely because the limiting factor in the solution accuracy is the low angular resolution being unable to match the flux surfaces to the boundary Fourier series. At higher angular resolutions, it can be seen that the FTOL parameter being too low limits the accuracy of the solution, as expected as it terminates the solver prematurely. The difference in the solutions found using $FTOL=1E-8$ and $FTOL=1E-12$ becomes larger as the angular resolution of the solution is increased.

\end{document}